\documentclass[twocolumn]{aastex631}
\usepackage{graphicx}	
\usepackage{amsmath}	

\usepackage{amssymb}	

\newcommand{\SiII}{Si~{\sc ii}}
\newcommand{\SII}{S~{\sc ii}}

\newcommand{\NaI}{Na~{\sc i}~D}
\newcommand{\SiIII}{Si~{\sc iii}}
\newcommand{\OI}{O~{\sc i}}

\newcommand{\CII}{C~{\sc ii}}
\newcommand{\FeIII}{Fe~{\sc iii}}

\newcommand{\NiII}{Ni~{\sc ii}}
\newcommand{\FeII}{Fe~{\sc ii}}
\newcommand{\CaII}{Ca~{\sc ii}}
\newcommand{\CaIII}{Ca~{\sc iii}}

\newcommand{\ld}{$\lambda$}

\shorttitle{SN 2024gy Delayed Detonation}
\shortauthors{Li et al.}
\graphicspath{{./}{24gy_fig/}}

\begin{document}

\title{SN 2024gy: Multi-epoch Spectroscopic Features Suggestive of Delayed Detonation in a Type Ia Supernova}

\author[0009-0003-3758-0598]{Liping Li} 
\affiliation{Yunnan Observatories, Chinese Academy of Sciences, Kunming 650216, China}
\affiliation{International Centre of Supernovae, Yunnan Key Laboratory, Kunming 650216, China}
\affiliation{Key Laboratory for the Structure and Evolution of Celestial Objects, Chinese Academy of Sciences, Kunming 650216, China}

\author[0009-0009-2664-8212]{Zhenyu Wang} 
\affiliation{Yunnan Observatories, Chinese Academy of Sciences, Kunming 650216, China}
\affiliation{School of Astronomy and Space Science, University of Chinese Academy of Sciences, Beijing 100049,1408, People's Republic of China}

\author[0009-0000-0314-6273]{Jialian Liu} 
\affiliation{Physics Department, Tsinghua University, Beijing 100084, China}

\author[0009-0002-7625-2653]{Yu Pan}
\affiliation{South-Western Institute for Astronomy Research (SWIFAR), Yunnan University, Kunming, Yunnan 650500, People's Republic of China}
\affiliation{Yunnan Key Laboratory of Survey Science, Yunnan University, Kunming, Yunnan 650500, People's Republic of China}

\author[0000-0003-3460-0103]{Alexei V. Filippenko}
\affiliation{Department of Astronomy, University of California, Berkeley, CA 94720-3411, USA}

\author[0000-0002-8296-2590]{Jujia Zhang*} 
\affiliation{Yunnan Observatories, Chinese Academy of Sciences, Kunming 650216, China}
\affiliation{International Centre of Supernovae, Yunnan Key Laboratory, Kunming 650216, China}
\affiliation{Key Laboratory for the Structure and Evolution of Celestial Objects, Chinese Academy of Sciences, Kunming 650216, China}
\email{* jujia@ynao.ac.cn}

\author[0000-0002-7334-2357]{Xiaofeng Wang} 
\affiliation{Physics Department, Tsinghua University, Beijing 100084, China}

\author[0000-0001-7225-2475]{Brajesh Kumar}
\affiliation{South-Western Institute for Astronomy Research (SWIFAR), Yunnan University, Kunming, Yunnan 650500, People's Republic of China}
\affiliation{Yunnan Key Laboratory of Survey Science, Yunnan University, Kunming, Yunnan 650500, People's Republic of China}

\author[0000-0002-6535-8500]{Yi Yang}
\affiliation{Physics Department, Tsinghua University, Beijing 100084, China}

\author[0000-0001-5955-2502]{Thomas G. Brink}
\author[0000-0002-2636-6508]{WeiKang Zheng}
\affiliation{Department of Astronomy, University of California, Berkeley, CA 94720-3411, USA}

\author[0000-0001-5316-2298]{Xiangcun Meng} 
\affiliation{Yunnan Observatories, Chinese Academy of Sciences, Kunming 650216, China}
\affiliation{International Centre of Supernovae, Yunnan Key Laboratory, Kunming 650216, China}
\affiliation{Key Laboratory for the Structure and Evolution of Celestial Objects, Chinese Academy of Sciences, Kunming 650216, China}

\author[0000-0002-1094-3817]{Lingzhi Wang} 
\affiliation{Chinese Academy of Sciences South America Center for Astronomy (CASSACA), National Astronomical Observatories, CAS, Beijing, China}
\affiliation{Departamento de Astronom{\'i}a, Universidad de Chile, Las Condes, 7591245 Santiago, Chile}

\author[0009-0005-2963-7245]{Zeyi Zhao} 
\affiliation{Yunnan Observatories, Chinese Academy of Sciences, Kunming 650216, China}
\affiliation{School of Astronomy and Space Science, University of Chinese Academy of Sciences, Beijing 100049,1408, People's Republic of China}

\author{Qian Zhai} 
\affiliation{Yunnan Observatories, Chinese Academy of Sciences, Kunming 650216, China}
\affiliation{Key Laboratory for the Structure and Evolution of Celestial Objects, Chinese Academy of Sciences, Kunming 650216, China}

\author[0000-0002-7714-493X]{Yongzhi Cai} 
\affiliation{Yunnan Observatories, Chinese Academy of Sciences, Kunming 650216, China}
\affiliation{International Centre of Supernovae, Yunnan Key Laboratory, Kunming 650216, China}
\affiliation{Key Laboratory for the Structure and Evolution of Celestial Objects, Chinese Academy of Sciences, Kunming 650216, China}

\author{Giuliano Pignata} 
\affil{Instituto de Alta Investigaci\'on, Universidad de Tarapac\'a, Casilla 7D, Arica, Chile}

\author[0009-0000-4068-1320]{Xinlei Chen}
\author[0009-0006-5847-9271]{Xingzhu Zou}
\author{Jiewei Zhao}
\author[0000-0003-0394-1298]{Xiangkun Liu}
\author[0000-0003-1295-2909]{Xiaowei Liu}
\affiliation{South-Western Institute for Astronomy Research (SWIFAR), Yunnan University, Kunming, Yunnan 650500, People's Republic of China}
\affiliation{Yunnan Key Laboratory of Survey Science, Yunnan University, Kunming, Yunnan 650500, People's Republic of China}

\author{Xinzhong Er} 
\affiliation{Tianjin Astrophysics Center, Tianjin Normal University, Tianjin, 300387, China}

\author[0000-0003-4254-2724]{A. Reguitti} 
\affiliation{INAF - Osservatorio Astronomico di Padova, Vicolo dell'Osservatorio 5, 35122 Padova, Italy}
\affiliation{INAF-Osservatorio Astronomico di Brera, Via E. Bianchi 46, 23807 Merate (LC), Italy}


\author[0000-0003-0427-8387]{R. Michael Rich}
\affiliation{Dept of Physics and Astronomy, University. of California, Los Angeles, 90095-1547}

\author[0000-0002-5376-3883]{Jon M. Rees}
\affiliation{UCO/Lick Observatory, Mount Hamilton, CA 95140}

\author[0000-0002-2908-9702]{Mark A. Croom}
\affiliation{NASA (retired), Hampton, VA 23681}

\author[0009-0001-3776-7868]{Osmin Caceres}
\affiliation{University of California, Los Angeles}

\author{K. Itagaki}
\affiliation{Itagaki Astronomical Observatory, Yamagata 990-2492, Japan}

\author[0000-0002-3231-1167]{Bo Wang**} 
\affiliation{Yunnan Observatories, Chinese Academy of Sciences, Kunming 650216, China}
\affiliation{International Centre of Supernovae, Yunnan Key Laboratory, Kunming 650216, China}
\affiliation{Key Laboratory for the Structure and Evolution of Celestial Objects, Chinese Academy of Sciences, Kunming 650216, China}
\email{** wangbo@ynao.ac.cn}

\author{Jinming Bai***} 
\affiliation{Yunnan Observatories, Chinese Academy of Sciences, Kunming 650216, China}
\affiliation{International Centre of Supernovae, Yunnan Key Laboratory, Kunming 650216, China}
\affiliation{Key Laboratory for the Structure and Evolution of Celestial Objects, Chinese Academy of Sciences, Kunming 650216, China}
\email{*** baijinming@ynao.ac.cn}

\begin{abstract}
We present photometric and spectroscopic observations of SN 2024gy, a Type Ia supernova (SN Ia) exhibiting high-velocity features (HVFs) in its early-time spectra. This SN reaches a peak $B$-band magnitude of $-19.25 \pm 0.29$ mag and subsequently declines by $\Delta m_{15}(B) \approx 1.12$ mag, consistent with the luminosity-width relation characteristic of normal SNe Ia. Based on the peak thermal luminosity of $(1.2 \pm 0.3) \times 10^{43}$ erg s$^{-1}$, we estimate that $0.57 \pm 0.14~\rm M_{\odot}$ of $^{56}$Ni was synthesized during the explosion. Our dense early spectral monitoring revealed significant velocity disparities within the ejecta. Notably, absorption features from the \CaII\ near-infrared triplet were observed at velocities exceeding 25,000 km s$^{-1}$, while the \SiII\, \ld 6355 line velocity at the same epoch was significantly lower at $\sim$ 16,000 km s$^{-1}$. This velocity disparity likely reflects distinct ionization states of intermediate-mass elements in the outermost layers. The prominent \CaII\, HVFs may originate from ionization suppression within the highest-velocity ejecta, potentially indicative of minimal hydrogen mixing in a delayed-detonation explosion scenario. Additionally, the Ni/Fe ratio derived from the nebular spectrum of SN 2024gy provides further support for this model. 
\end{abstract}



\section{Introduction} \label{sec:1}

Type Ia supernovae (SNe Ia) result from thermonuclear explosions of carbon-oxygen white dwarfs in binary systems \citep[e.g.,][]{1997Sci...276.1378N,2000ARA&A..38..191H,2012NewAR..56..122W,2014ARA&A..52..107M}. As the premier standardizable candles with precisely calibrated luminosities \citep{1993ApJ...413L.105P}, they revolutionized modern cosmology by enabling the discovery of cosmic acceleration \citep{1998AJ....116.1009R,1999ApJ...517..565P}.

However, persistent discrepancies in measurements of the Hubble constant \citep[e.g.,][]{2022ApJ...934L...7R} highlight the need for tighter observational constraints on progenitor systems. The single-degenerate (SD) scenario suggests a white dwarf accreting material from a nondegenerate companion until approaching the Chandrasekhar mass limit \citep[$\sim1.4~{\rm M}_\odot$][]{1973ApJ...186.1007W,1984ApJ...286..644N}. This framework is supported by the detection of circumstellar material (CSM) in some SNe Ia, as observed in SN 2017cbv \citep{2017ApJ...845L..11H,2020ApJ...904...14W}. 
Indeed, there is certainly a subtype of SNe~Ia surrounded by CSM, called SN~Ia-CSM, such as SN 2002ic \citep[][]{2003Natur.424..651H}, PTF 11kx \citep[][]{2012Sci...337..942D} and SN 2018evt \citep[][]{2023MNRAS.519.1618Y,2024NatAs...8..504W}.
Alternatively, the double-degenerate (DD) model invokes mergers of two white dwarfs \citep{1984ApJS...54..335I,1984ApJ...277..355W}.


Observationally, most SNe Ia exhibit homogeneous photometric properties, which are conventionally referred to as ``normal'' SNe Ia. 
These events are characterized by a well-documented empirical relationship between light-curve width (parameterized by $\Delta m_{15}(B)$) and peak luminosity, historically termed the ``Phillips relation'' (or the ``width--luminosity relation,'' WLR; \citealt{1993ApJ...413L.105P,1999AJ....118.1766P}). 
%
However, optical spectra of SNe Ia display significant diversity \citep{1997ARA&A..35..309F}.



Despite the unclear origin of the spectroscopic diversity, more refined classification schemes have been developed to correlate spectroscopic features with possible underlying physical parameters.
These include systems based on the velocity gradients \citep[][]{2005ApJ...623.1011B}, the pseudo-equivalent width (pEW) ratio of \SiII\ \citep[][]{2006PASP..118..560B}, or the velocity of photospheric \SiII\ at maximum light \citep[][]{2009ApJ...697..380W}.
Those schemes may have captured the differences on the ejecta in the explosion, such as the nucleosynthesis, temperature, ionization state, and kinematics.

The spectroscopic diversity of SNe Ia extends beyond the well-characterized photospheric velocity (PV) components to the high-velocity features (HVFs), blueshifted absorption lines offset by 6000--13,000 km s$^{-1}$ from the PV \citep[e.g.,][]{2005MNRAS.357..200M,2016ApJ...817..114Z}. 
HVFs were initially identified in \CaII\,H$\&$K and the near-infrared triplet (IRT) \citep{1999ApJ...525..881H}. Such features are characterized by multi-element emission dominance during their early phases ($t \leq -5$  days) in approximately 95\% of \CaII\, IRT cases \citep{2014MNRAS.444.3258M}, manifesting through \SiII\, \ld6355 absorption in $\sim 20$\% of SNe~Ia \citep{2014MNRAS.437..338C}. 
Occasional detections have also been reported in \SiIII, \SII, and \FeII\, line profiles \citep{2013ApJ...777...40M}.

These HVFs rapidly decay after maximum light as the PV components become more prominent \citep[e.g.,][]{2015ApJS..220...20Z}. 
Spectropolarimetric studies have revealed that HVFs form in asymmetric ejecta regions, which are distinct from the more spherical geometry of the PV features \citep{2006ApJ...653..490W,2009A&A...508..229P}.
Systematic analyses of large samples ($N \geq 264$) uncover anticorrelations between \CaII\,IRT HVF strength and both $\Delta m_{15}(B)$ and $v_{\text{Si}}$, where rapidly declining events lack HVFs \citep{2014MNRAS.444.3258M,2014MNRAS.437..338C}. 

Conversely, \SiII\, \ld6355 HVFs preferentially occur in high-$v_{\text{Si}}$ SNe Ia with redder peak colors and absent \CII\,  absorption \citep{2015MNRAS.451.1973S}. 
Additionally, different-strength HVFs have been proposed to be linked to different stellar populations \citep{2019ApJ...886...58M}. The physical origin of HVFs remains debated. Current evidence suggests that HVFs may originate from at least one of the following mechanisms: density enhancement, abundance enhancement, or ionization effects occurring in a high-velocity region \citep[e.g.,][]{2005MNRAS.357..200M}. Intensive spectroscopic monitoring and analysis of SNe Ia exhibiting early-phase HVFs represent a crucial approach to resolving these controversies.

In this paper, we present a detailed analysis of SN~2024gy, an SN Ia characterized by exceptionally high-velocity \CaII\, IRT HVFs. 
The observations are described in Section \ref{sec:2}.
Sections \ref{sec:3} and \ref{sec:4} respectively present
the light-curve and spectral analyses.
We discuss the luminosity,  the possible $^{56}$Ni mass, and the HVFs in Section \ref{sec:5}.
Our main conclusions are summarized in Section \ref{sec:6}.

\section{observations and data reduction} \label{sec:2}

\begin{figure}
    \centering
    \includegraphics[width=\linewidth]{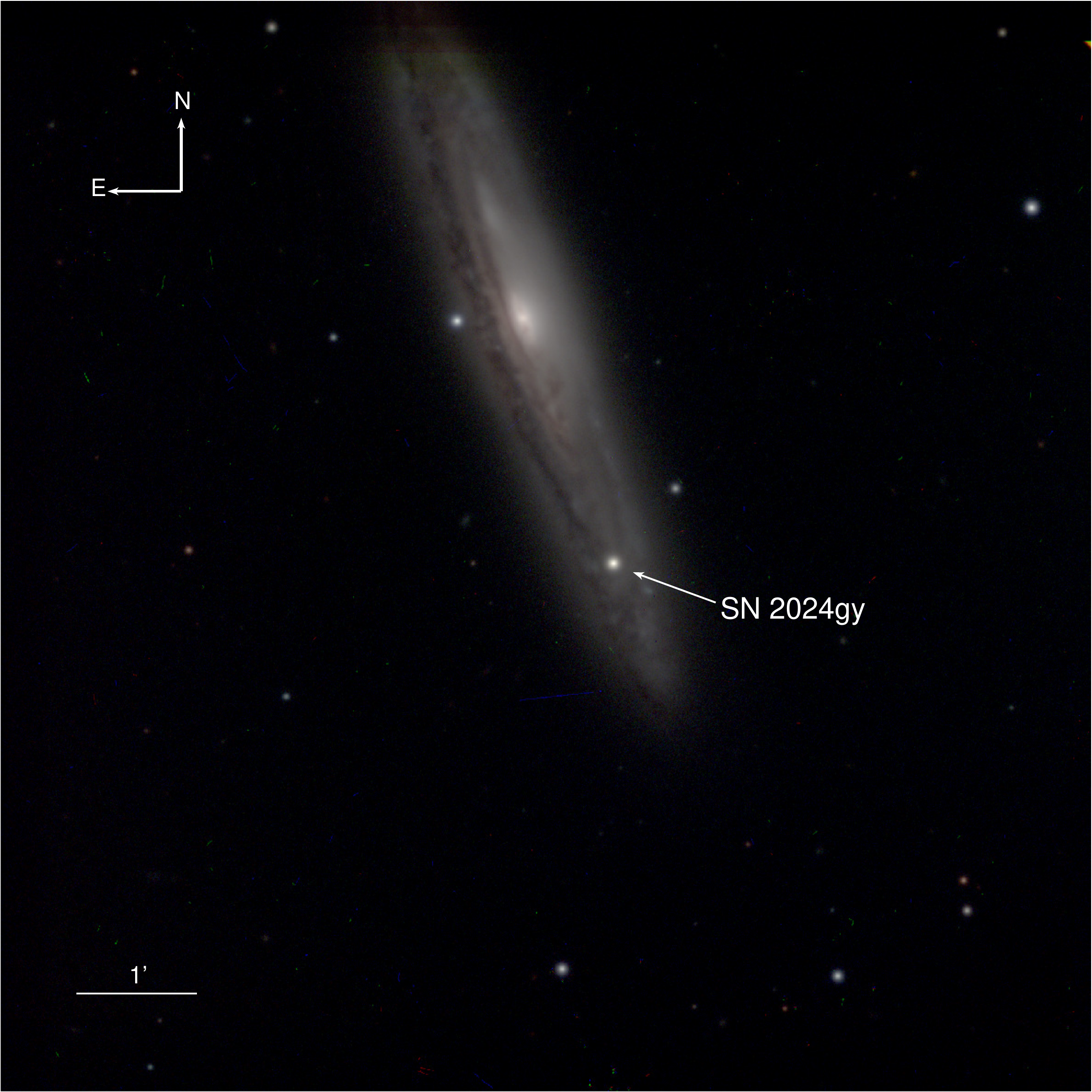}
    \caption{Lijiang 2.4\,m telescope color composite (\textit{B/g/r}) image of SN 2024gy and its host galaxy NGC 4216.
    }
    \label{fig:loc}
\end{figure}

SN 2024gy was discovered by Koichi Itagaki on 2024-01-04.678 (UTC dates and times are used throughout this paper) in the clear filter at a brightness of 16.3 mag \citep[][]{2024TNSTR..39....1I}. 
Its J2000 coordinates are $\alpha =$ {12$^{\rm hr}$}{15$^{\rm m}$}{51$\fs$290}, $\delta =$ +{13$\degr$}{06$\arcmin$}{56$\farcs$12}, and it is located in the spiral galaxy NGC~4216 as shown in Figure \ref{fig:loc}. 
It was classified as a young Type Ia SN by the Global SN Project \citep{2024TNSCR..44....1N}, with extremely HVFs of the \CaII\, IRT.

\subsection{Photometry} \label{subsec:photometry}

Optical photometry of SN 2024gy in the \textit{uBVgri} bands was obtained by the Lijiang 2.4\,m telescope \citep[LJT+YFOSC;][]{2015RAA....15..918F,2019RAA....19..149W} from modified Julian Dates (MJD) 60314.84 to 60443.73 ($t\approx -15$  to $+114$ d; hereafter, $t$ represents the time relative to   $B$-band maximum brightness given in Section \ref{subsec:LC}).
The 0.8\,m Tsinghua-NAOC telescope \citep[TNT;][]{2012RAA....12.1585H} also monitored SN~2024gy in the \textit{BVgri} bands from MJD 60314.87 to 60417.55. Aperture or template-subtracted photometry was performed using the AutoPhOT \citep{2022A&A...667A..62B} pipeline for LJT and TNT.

Furthermore, SN~2024gy was monitored by the 1.6\,m Multi-channel Photometric Survey Telescope \citep[Mephisto;][]{Yuan-2020-Mephisto, 2024ApJ...971L...2C}, which is under the commissioning phase. Observations were performed between MJD~60314.9 and 60645.9 ($t\approx -15$ to $+316$~d). The $uvgriz$ Mephisto filters have wavelength coverage of 320–-365 nm ($u$ band), 365–-405 nm ($v$), 480–-580 nm ($g$), 580–-680 nm ($r$), 775–-900 nm ($i$), and 900–-1050 nm ($z$), and the corresponding central wavelengths are 345, 385, 529, 628, 835, and 944\,nm, respectively \citep{2024ApJ...971L...2C, Yuan-PeiY, Guowang2025D, Zou-2025}. The images were obtained simultaneously in the $ugi$ bands or in the $vrz$ bands. 

As the SN is located inside the host (see Fig.~\ref{fig:loc}), image subtraction was implemented on all images to remove the galaxy  contamination. The template frames acquired on 2025 March 27 ($\sim 447$ days after discovery) were used as reference. 
The Mephisto magnitudes were calibrated employing the Synthetic Photometry Method and by utilizing the Gaia BP/RP low-resolution XP spectra obtained from the Gaia satellite \citep{2023A&A...674A...2D, 2023A&A...674A...3M}. The detailed procedure is presented by \citet{2024ApJ...971L...2C} and \citet{Zou-2025}. Photometric calibration uncertainties were better than 0.03 mag, 0.01 mag, and 0.005 mag in the $u$, $v$, and $griz$ bands, respectively.


Some optical and near-infrared photometric observations spanning from $-14.3$ d to +35.7 d were obtained with REM \citep[][]{2004SPIE.5492.1613C,2004SPIE.5492.1590Z}, a 60\,cm rapid-response telescope located at the ESO La Silla Observatory in Chile. REM is equipped with two instruments: REMIR, an infrared imaging camera \citep[][]{2004SPIE.5492.1602C}, and ROS2, a visible-light camera capable of simultaneous imaging in four Sloan filters \citep[\textit{g}, \textit{r}, \textit{i}, \textit{z};][]{2004SPIE.5492..689T}. Thanks to a dichroic-based optical design, REM can acquire five images simultaneously -- covering the $griz$ bands and an additional infrared band.
Additionally, five epochs of $BVgri$ photometry were obtained with the robotic 67/92\,cm Schmidt telescope located on Mount Ekar in Asiago, Italy.
Aperture photometry of SN 2024gy and the comparison stars was performed after template subtraction using standard \texttt{IRAF}\footnote{\texttt{IRAF}, the Image Reduction and Analysis Facility, is distributed by the National Optical Astronomy Observatory, which is operated by the Association of Universities for Research in Astronomy (AURA), Inc. under cooperative agreement with the U.S. National Science Foundation (NSF).} routines. 
The \textit{griz}-band and \textit{H}-band templates were taken from the Sloan Digital Sky Survey and the Two Micron All-Sky Survey.
The template-subtracted procedures were performed using the SFFT\footnote{\url{https://github.com/thomasvrussell/sfft}} \citep[Saccadic Fast Fourier Transform;][]{2022ApJ...936..157H} software.

The Zwicky Transient Facility (ZTF) started  monitoring SN 2024gy (named ZTF24aaaiocv) on 2024-01-12 in the \textit{g} and \textit{r} bands.
For a data complement, we download the public data\footnote{\url{https://lasair-ztf.lsst.ac.uk/objects/ZTF24aaaiocv/}} of ZTF from the start to 2024-05-16.

The photometric data of LJT, TNT, REM, Schmidt, and ZTF are listed in Table \ref{app_a:photo}. The Mephisto data are given in Table \ref{app_a:M_photo}.

\subsection{Spectroscopy}
Low-resolution spectra of SN~2024gy were obtained with the Lijiang 2.4\,m telescope \citep[LJT+YFOSC][]{Wang-2019RAA}, the Xinglong 2.16\,m telescope \citep[XLT+BFOSC;][]{2016PASP..128j5004Z}, the Kast double spectrograph on the 3\,m Shane telescope at Lick Observatory \citep{miller1994lick}, and the low resolution imaging spectrograph (LRIS) on the 10\,m Keck-I telescope at Keck observatory \citep{1995PASP..107..375O}.
There are 4 mid-resolution spactra obtained by LJT.
Information for all spectroscopic observations is listed in Table \ref{app_a:spec}.

All spectra obtained by LJT and XLT were reduced using standard IRAF routines. The spectra were corrected for the local atmospheric extinction and calibrated with  spectrophotometric standard stars observed at a similar airmass on the same night. The telluric lines were also removed.
The Kast spectra utilized the $2^{''}$ slit, the D57 dichroic, the 600/4310 grism, and either the 300/7500, 600/7500, or 830/8460 grating.  To minimize slit losses caused by atmospheric dispersion \citep[][]{1982PASP...94..715F}, the slit was oriented at or near the parallactic angle. The data were reduced following standard techniques for CCD processing and spectrum extraction \citep[][]{2012MNRAS.425.1789S} utilizing IRAF routines and custom Python and IDL codes\footnote{\url{ https://github.com/ishivvers/TheKastShiv}}.  Low-order polynomial fits to comparison-lamp spectra were used to calibrate the wavelength scale, and small adjustments derived from night-sky lines in the target frames were applied. The spectra were flux calibrated using observations of appropriate spectrophotometric standard stars observed on the same night, at similar airmasses, and with an identical instrument configuration.
The LRIS spectra utilized the $1^{''}$ slit, the D560 dichroic, the 600/4000 grism, and the 400/8500 grating.  This instrument configuration produced a combined wavelength range of $\sim$3200–10,200 \AA, and a spectral resolving power of R$\approx$ 900.  The LRIS spectra were reduced with the LPipe data reduction pipeline \citep[][]{2019PASP..131h4503P}.

\section{photometry analysis} \label{sec:3}
\subsection{Light Curves} \label{subsec:LC}

\begin{figure*}
    \centering
    \includegraphics[width=0.8\textwidth]{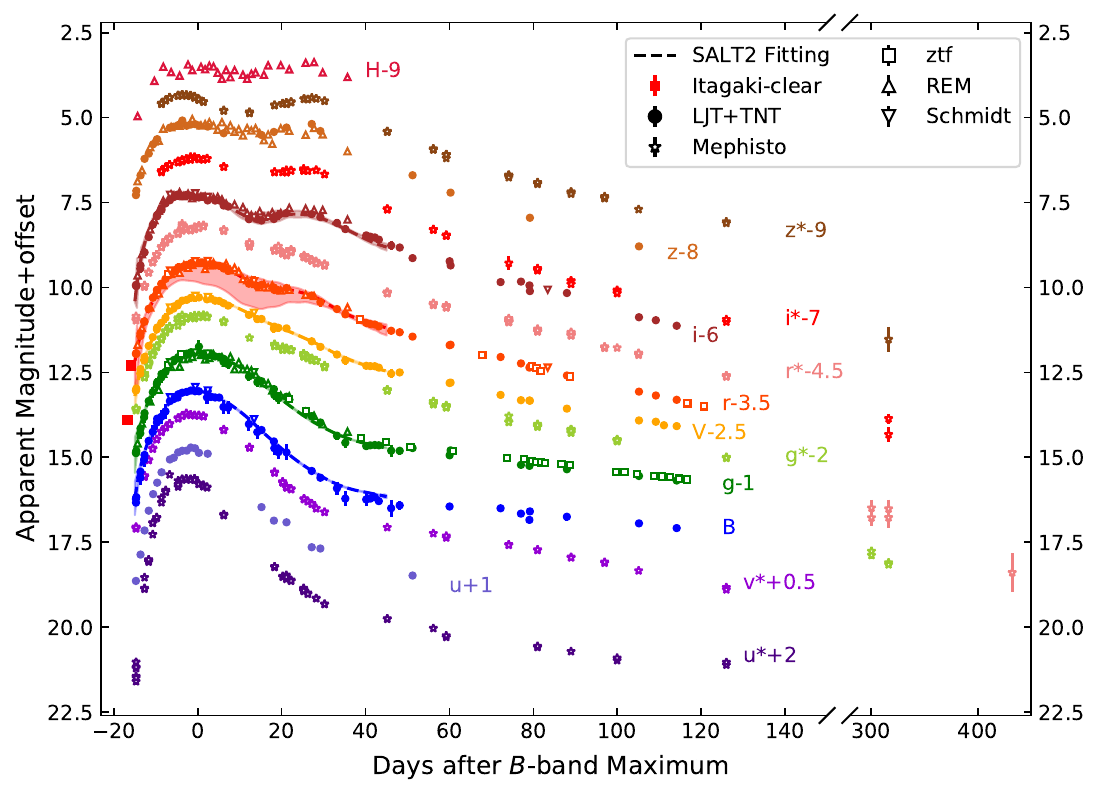}
    \caption{Optical light curves of SN~2024gy. Phase is relative to the time of $B$-band maximum light. Data in different filters are shown in different colors and shifted vertically for better display. Mephisto bands are marked with asterisks to distinguish them from SDSS bands. The dashed lines are the best fits of the SALT2 model. Regions with different colors are the  1$\sigma$ confidence interval for the best fits. 
    }
    \label{fig:LC1}
\end{figure*}

Figure \ref{fig:LC1} shows the optical and near-infrared light curves of SN~2024gy.
The Mephisto bands are marked with asterisks to distinguish them from SDSS bands.
The optical observations cover the phases from $t \approx -15$ d to $t \approx 433$ d after \textit{B}-band maximum light.
The shapes of the light curves are similar to those of typical SNe Ia. 
A shoulder appears after the first peak ($t\approx +25$ d) in the $r$/$r^*$ bands. The secondary maximum can be identified in the $i/i^*/z/z^*/H$ bands, and they reach their primary peaks earlier than in the \textit{B} band.
Through polynomial fitting of the light curves of SN~2024gy near peak brightness, we determine that the SN reaches 
$B_{\max} = 13.03 \pm 0.02$ mag on MJD $60328.8  \pm 0.3$. The corresponding decline rate is $\Delta m_{15}(B) = 1.16 \pm 0.04$ mag.

We also model the optical light curves of SN 2024gy using the SALT2 template \citep[the spectral adaptive light-curve template;][]{2007A&A...466...11G}.
The best-fit results are shown in Figure~\ref{fig:LC1} (dashed lines and colored confidence regions), with the corresponding best parameters of $x_1=-0.20\pm0.05$, $c=0.20\pm0.02$, and $B_{\max} = 13.05 \pm 0.02$ mag. 
According to \citet{2007A&A...466...11G}, the $\Delta m_{15}(B)$ from the SALT2 fitting can be roughly transformed with an equation:
\begin{equation*}
\Delta m_{15}(B) = 1.09-0.161x_1+0.013x_{1}^{2}-0.00130x_1^3.
\end{equation*}
With the best-fit $x_1$, the SALT2 fitting yields $\Delta m_{15}(B) = 1.12 \pm 0.01$ mag for SN 2024gy. 
These measurements agree with the polynomial fit results. 
We adopt MJD $60329.5 \pm 0.3$ as the time of \textit{B}-band maximum light in the analyses throughout this paper.

Late-time photometry from Mephisto (300 d to 433 d) reveals an $r^*$-band decline rate of $\sim 1.2$ mag per 100 days, faster than the rate of $^{56}Co$ decay (0.98 mag per 100 days). 

\begin{deluxetable*}{lcccc}
\tablecaption{The information of the SNe Ia used for comparison\label{tab:compared_SNs}}
\tablewidth{0pt}
\tablehead{
\colhead{Object} & \colhead{$\Delta m_{15}(B)$} & \colhead{$(B_{\rm max}-V_{\rm max})_0$\tablenotemark{$\dagger$}} & \colhead{$E(B-V)_{\rm host}$} & \colhead{Ref.}\\
\colhead{}  & \colhead{mag} & \colhead{mag} & \colhead{mag} & \colhead{}
}
\startdata
SN 2024gy   & $1.12\pm0.02$ & $-0.15\pm0.12$ & $0.38\pm0.10$ &  this work \\
SN 2011fe  & $1.18\pm0.03$ & $-0.05\pm0.05$ & $0.03\pm0.05$ & (1)(2) \\
SN 2017erp & $1.05\pm0.06$  & $-0.17\pm0.03$ & $0.18\pm0.03$ &  (3)\\
SN 2019np  & $1.04\pm0.04$  & $-0.05\pm0.05$ & $0.10\pm0.04$ &  (4)\\
SN 2021fxy  & $1.05\pm0.06$  & $-0.05\pm0.06$ & $0.02\pm0.06$ & (5) \\
\enddata
\tablenotetext{\dagger}{This column is the intrinsic color at \textit{B}-band maximum. The values are calculated using the dereddened color curves shown in Figure \ref{fig:CC}. The uncertainties of $E(B-V)$ are also involved.}
\tablerefs{(1) \cite{2016ApJ...820...67Z}, (2) \cite{2013CoSka..43...94T},(3) \cite{2019ApJ...877..152B}, (4) \cite{2022MNRAS.514.3541S}, (5) \cite{2023MNRAS.522.3481D}.}
\end{deluxetable*}

\begin{figure*}
\centering
\includegraphics[width=0.85\textwidth]{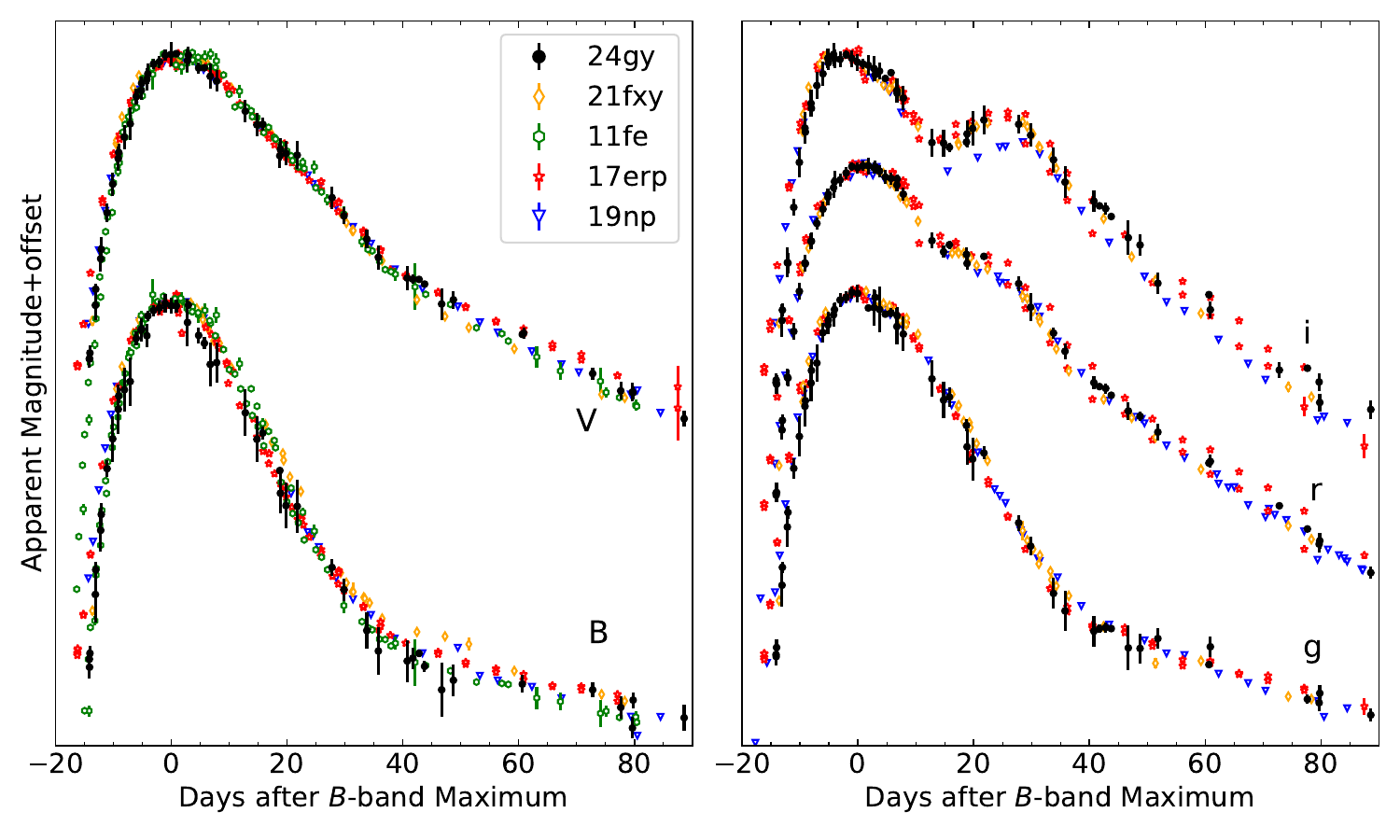}
\caption{Comparison of the optical light curves of SN~2024gy with those of other well-observed SNe Ia, including SN~2011fe, SN~2019np, SN~2017erp, and SN~2021fxy. The light curves of the compared SNe are normalized to match the peak magnitudes of SN~2024gy in the respective bands.}
    \label{fig:LCcom}
\end{figure*}

Figure \ref{fig:LCcom} compares the light curves of SN 2024gy with those of several well-sampled normal SNe Ia exhibiting similar decline rates, including SN 2011fe, SN 2017erp,  SN 2019np, and SN 2021fxy.
The light-curve information of those SNe are listed in Table \ref{tab:compared_SNs}.
The light-curve morphology of SN 2024gy is similar to that of SN 2021fxy in all bands and to that of SN 2011fe in the \textit{BV} bands.


\subsection{Reddening} \label{subsec:reddening}
SN 2024gy is located on a spiral arm of the nearby edge-on galaxy NGC 4216, at an angular distance of $\sim2\arcmin$ from the galactic center, as shown in Figure \ref{fig:loc}.
The redshift of the host galaxy is $z=0.000437$ according to the NASA/IPAC Extragalactic Database (NED\footnote{\url{https://ned.ipac.caltech.edu/}}). 
This value is adopted for all analyses presented in this work.



The Galactic extinction toward SN 2024gy is estimated as $A_{V}^{\rm Gal} = 0.087$ mag according to the dust map derived by \citet{2011ApJ...737..103S}. 
Adopting the extinction law with a
total-to-selective extinction ratio
$R_{V}^{\rm Gal} = 3.1$ \citep[e.g.,][]{1989ApJ...345..245C}, the reddening due to the Milky Way is $E(B - V)_{\rm Gal} = 0.028$ mag.

The total reddening can be estimated using the Lira-Phillips (LP) relation \citep{1999AJ....118.1766P}, which is shown in Figure \ref{fig:CC}.
A reddening of $E(B-V)=0.42 \pm 0.10$ mag is given by matching the $B-V$ color curve ($30\, {\rm d}\leq t \leq 90\, {\rm d}$) to the LP relation.
However, the data are not well fitted to the LP relation.
Alternatively, using Equation (7) of \cite{1999AJ....118.1766P} with the values of $\Delta m_{15}$ and $B_{\rm max}-V_{\rm max}$ (the pseudocolor at maximum light), we get a similar reddening of $E(B-V)=0.40 \pm 0.03$ mag.
These estimates yield a similar result, indicating a relatively large reddening for SN 2024gy.
Given the large uncertainties in the current estimates, the results require validation through alternative methods.

The SALT2 fitting result gives a color parameter $c$ of $\sim0.2$, also suggesting a relatively large reddening for SN 2024gy.
The EBV\_model fitting in SNooPy \citep{2011AJ....141...19B} gives a host reddening of $E(B-V)_{\rm host}=0.28\pm0.07$ mag.
The color\_model fitting gives a different result of $E(B-V)_{\rm host}=0.40\pm0.07$ mag with $R_V^{\rm host}=1.07\pm0.20$, yielding a reddening similar to that of the LP method.
However, the $R_V^{\rm host}$ is smaller than the regular assumption of $R_V=3.1$, which should be checked by other light-curve-fitting methods. 

We use SUGAR \citep{2020A&A...636A..46L} and MLCS2k2 \citep{2007ApJ...659..122J} in the SNCosmo\footnote{\url{https://sncosmo.readthedocs.io}} \citep{2022zndo....592747B} frame to fit the light curves.
The SUGAR fitting yields an extinction of $A_{V}=0.662\pm0.018$ mag.
Considering $A_{V}^{\rm Gal}$ = 0.087 mag and $E(B-V)_{\rm host}=0.38\pm0.07$ mag, $R_V^{\rm host}=1.5\pm0.1$ for the host galaxy can be obtained.
We then fix the host $R_V$ to 1.5 and perform the MLCS2k2 fitting, resulting in a reasonable reddening of $E(B-V)_{\rm host}=0.38\pm0.10$ mag.
Nevertheless, a color\_model fitting in SNooPy with fixed $R_V^{\rm host}$ (1.5) still gives a reddening of $E(B-V)_{\rm host}=0.38\pm0.07$ mag.

In summary, we use different methods to estimate the reddening for SN 2024gy.
The results are listed in Table~\ref{tab:reddening}. 
Most of the results converge to a consistent $E(B-V)_{\rm host}$ of $\sim0.38$ mag and a consistent $A_{V}$ of $\sim 0.7$ mag.
Therefore, we adopt $E(B-V)_{\rm host}=0.38\pm0.10$ mag with $R_V^{\rm host}=1.5$ for SN 2024gy in the following analyses, while the corresponding Galactic reddening is $E(B-V)_{\rm Gal}=0.028$ mag with $R_V^{\rm Gal}=3.1$. 


\begin{deluxetable*}{lcccc}
\tablecaption{Reddening Estimate of SN 2024gy\label{tab:reddening}}
\tablewidth{0pt}
\tablehead{
\colhead{Method} & \colhead{$E(B-V)$} & \colhead{$E(B-V)_{\rm host}$} & \colhead{$A_V$} & \colhead{$R_V^{\rm host}$}\\
\colhead{} & \colhead{mag} & \colhead{mag} & \colhead{mag} & \colhead{}
}
\startdata
LP relation & $0.42\pm0.10$  & --- & --- & ---  \\
peak color & $0.40\pm0.03$  & --- & --- & --- \\
SNooPy EBV\_model & --- & $0.28\pm0.07$  & --- & 3.1\tablenotemark{$\star$} \\
SNooPy color\_model & --- & $0.40\pm0.07$  & --- & $1.07\pm0.20$  \\
SNooPy color\_model & --- & $0.38\pm0.07$  & --- & 1.5\tablenotemark{$\star$}  \\
SUGAR & --- &  --- & $0.662\pm0.018$ & --- \\
MLCS2k2 & --- &  $0.38\pm0.10$ & --- & 1.5\tablenotemark{$\star$} \\
\enddata
\tablenotetext{\star}{These parameters are fixed in the fitting.}
\tablecomments{The values shown are the parameters in the corresponding estimation method. See the text for the derived reddening and other details.}
\end{deluxetable*}

\subsection{Color Curves}
\begin{figure*}
    \centering
    \includegraphics[width=0.49\linewidth]{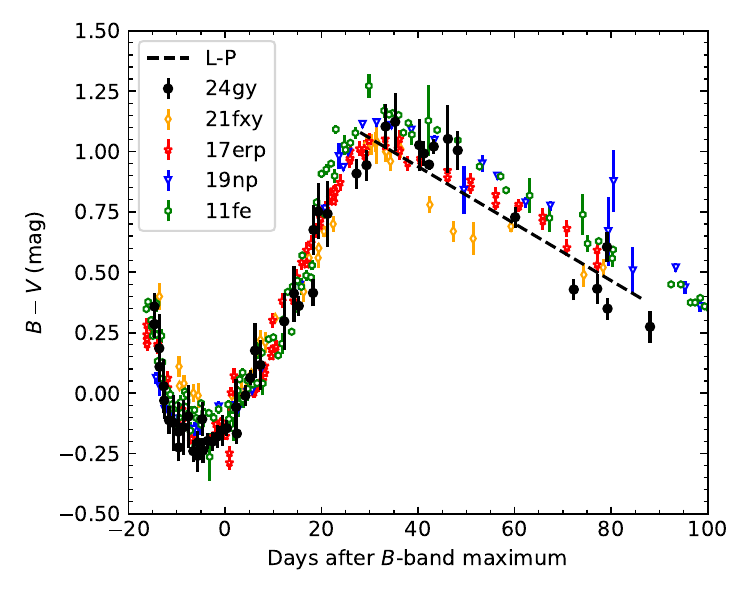}
    \includegraphics[width=0.49\linewidth]{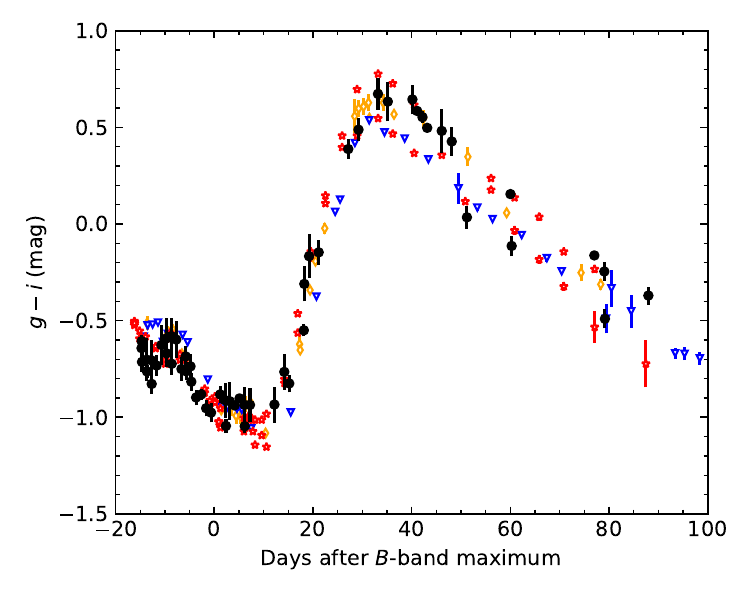}
    \caption{The $B-V$ and $g-i$ color curves of SN 2024gy (black circles),  compared with those of SNe~2011fe, 2017erp, 2019np, and 2021fxy (various markers in different colors). The dashed line is the LP relation. The color curves are all dereddened using the corresponding reddening listed in Table \ref{tab:compared_SNs}.}
    \label{fig:CC}
\end{figure*}

The $B-V$ and $g-i$ color curves of SN 2024gy are displayed in Figure \ref{fig:CC}, compared with those of other SNe Ia mentioned in Section \ref{subsec:LC}.
The reddening of $E(B-V)=0.40$ mag is adopted for SN 2024gy in the dereddening procedure.
One can see that the dereddened color curves of SN 2024gy match well those of the other normal SNe Ia.

The $B-V$ color curve of SN 2024gy monotonically becomes bluer at early phases. After $t\approx -5$ d, it turns red until $t\approx 30$ d.
The color curve then becomes bluer again, with a slightly steeper slope compared to the LP relation.
In general, the \textit{B-V} color evolution of SN 2024gy is consistent with that of the normal SNe Ia.
However, the intrinsic \textit{B-V} colors at maximum light for SN 2024gy and SN 2017erp seem to be bluer than that of the others (see Table \ref{tab:compared_SNs}).
In spectroscopic observation, the bluer colors may be caused by the different absorption of \FeII/\FeIII~ at maximum light.
The physical mechanism for this behavior is unclear.
It might be related to the temperature or the high-velocity ejecta in the explosion, such as the significant HVFs shown in SN 2024gy.
Nevertheless, the higher velocity ejecta yields a redder $B-V$ color \citep[e.g.,][]{2011ApJ...729...55F}, which is not consistent with the observation for SN 2024gy.

The $g-i$ color curve has a similar evolution. The differences are the locations of the two turning points. The first is located at $t\approx +10$ d, while the second is located at $t\approx +35$ d. Nevertheless, the dereddened $g-i$ color of SN~2024gy matches well with the other events, suggesting that the true reddening is estimated correctly. 

\section{Spectral analysis} \label{sec:4}
\begin{figure}
    \centering
    \includegraphics[width=1\columnwidth]{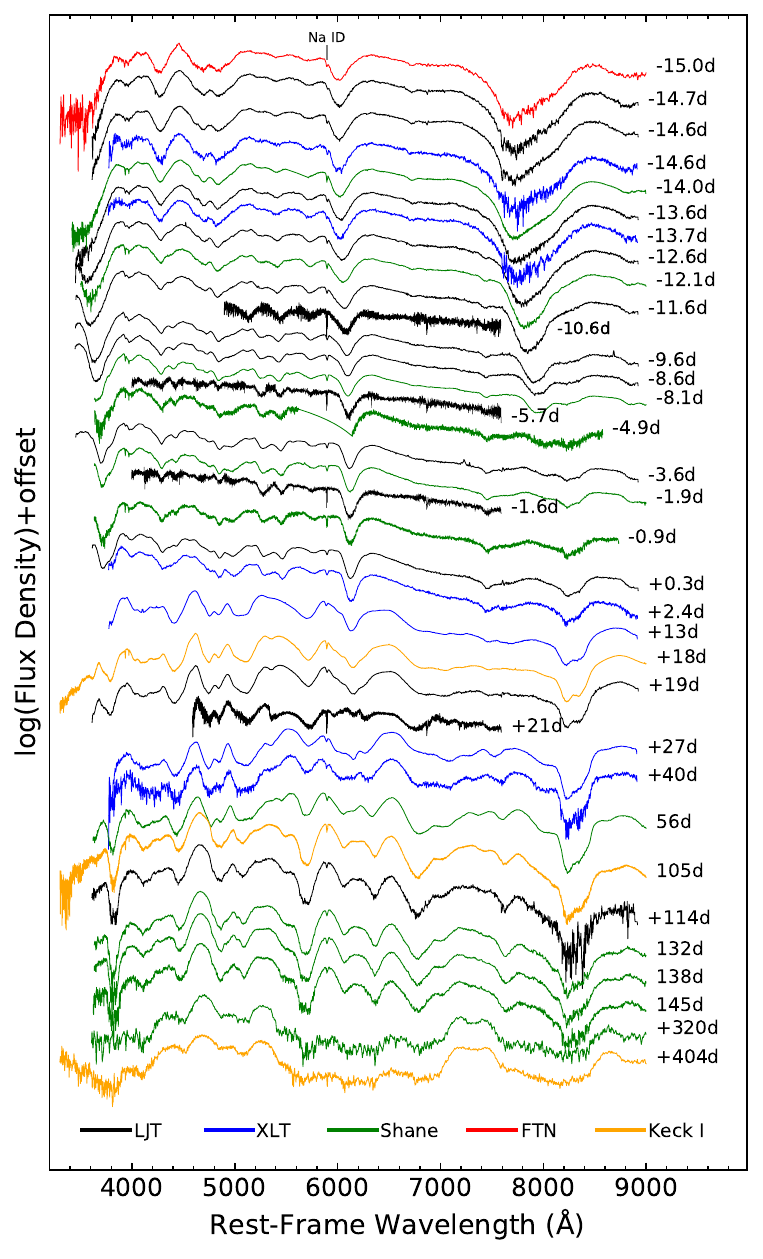}
    \caption{Optical spectral evolution of SN 2024gy. The spectra have been corrected for the redshift of the host galaxy and telluric lines. The numbers on the right-hand side mark the epochs of the spectra in days after the time of $B$-band maximum light. The position of the \NaI\,  is marked. The spectra in this figure is available as the Data behind the Figure.
    }
    \label{fig:spec}
\end{figure}

\subsection{Temporal Evolution of the Optical Spectra} \label{spectra}
The optical spectral evolution of SN 2024gy is displayed in Figure \ref{fig:spec}. There are a total of 35 optical spectra, spanning from $-15$ d to +404 d with respect to the time of \textit{B}-band maximum light. 
We also plot the public classification spectrum obtained with the Faulkes Telescope North (FTN).

The spectral evolution of SN 2024gy is similar to that of normal SNe Ia.
For instance, near the time of maximum light, spectra of SN 2024gy exhibit prominent absorption lines of intermediate-mass elements (IMEs; e.g., \SiII\, \ld6355, \CaII\ H\&K, \CaII\, IRT) and ionized IGEs. As SN 2024gy transitioned into the early nebular phase, the absorption features of IGEs 
dominated the spectra.
Notably, distinct sodium absorption features (\NaI\, \ld5889.95, 5895.92) are observable throughout the entire spectral evolution of SN~2024gy, which suggests that it may suffer significant extinction, as discussed in  Section \ref{subsec:reddening}.

\begin{figure*}
    \centering
    \includegraphics[width=0.95\textwidth]{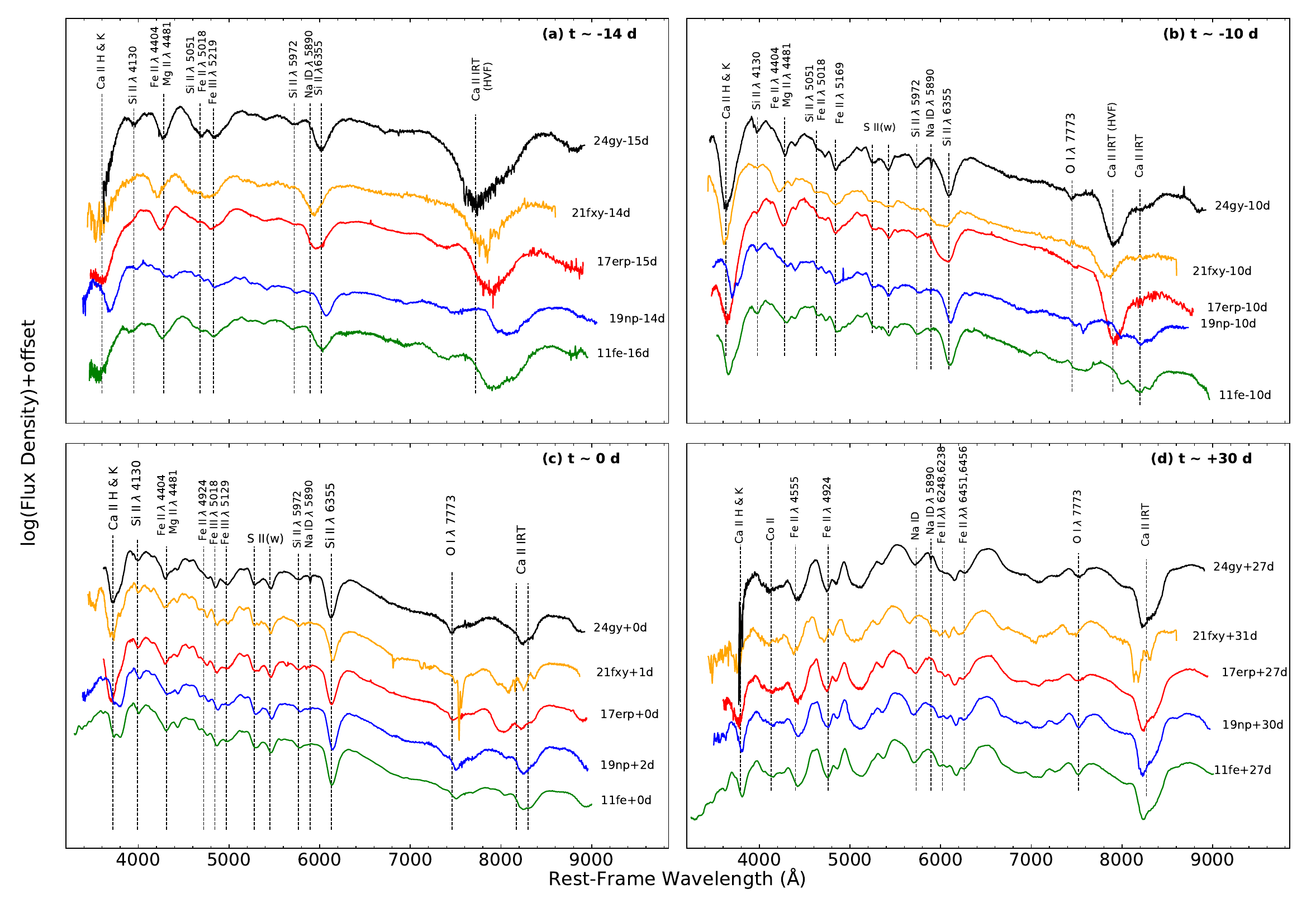}
    \caption{Spectral comparison of SN 2024gy (black) with other well-sampled SNe~Ia at typical phases (i.e., $t \approx -15$ d, $-10$ d, 0~d, and +27~d relative to \textit{B}-band maximum). The spectra marked in different colors are the comparable-phase spectra of SN~2011fe, SN 2019np, SN 2017erp, and SN 2021fxy. All spectra have been corrected for host-galaxy redshifts.
    The vertical dashed lines represent the wavelengths of the corresponding absorption lines of SN 2024gy. 
    }
    \label{fig:spcom}
\end{figure*}

Figure \ref{fig:spcom} shows comparative spectroscopy of SN~2024gy alongside normal SNe Ia with similar decline rates at four specific periods ($t \approx -15$, $-7$, 0, $+27$ d relative to  \textit{B}-band maximum). 
Panel (a) in Figure \ref{fig:spcom} presents the $t \approx -14$ d spectra, where all objects show absorption features of IMEs, though with notable variations in both line velocities and absorption depths. Similar to SNe 2021fxy and 2017erp, SN 2024gy displays prominent \CaII\,IRT HVFs, with the highest velocities measured in our sample. The \CaII\,IRT profiles already reveal clear photospheric components in SNe 2024gy, 2019np, and 2011fe at this epoch.

The velocity of \SiII\, \ld6355 absorption in SN 2024gy is smaller than that in SN 2021fxy and SN 2017erp, but larger than that in SN 2019np and SN~2011fe.
There could be an HVF of \SiII\, \ld6355 in the spectrum of SN 2024gy, but it is mixed with the absorption of \NaI\, \ld5892.
The EWs of the \SiII\, \ld4130, \SiII\, \ld5051 and \SiII\, \ld5972 absorption features in SN~2024gy spectra are significantly larger than those in the other samples.
In SN 2021fxy and SN 2017erp, \SiII\, lines are barely visible.
These discrepancies indicate different temperatures or different abundances of IMEs in the explosions.
The similar decline rates and intrinsic colors (listed in Table \ref{tab:compared_SNs}) at \textit{B}-band maximum for those SNe indicate the possible comparable photospheric temperatures.
More IMEs are potentially produced during the explosion of SN 2024gy.


By $t \approx -10$ d (panel (b) in Figure \ref{fig:spcom}), the ``W''-shaped \SII\, absorption features begin to become apparent. 
At this point, the absorption of \CaII\, IRT is still dominated by HVFs in SNe 2024gy, 2021fxy, and 2017erp, but there are clearly detached PV features of \CaII\, IRTs appearing in SN 2024gy.

Panel (c) in Figure \ref{fig:spcom} compares spectra around maximum brightness. During this phase, the spectra are characterized by  prominent absorption features of \CaII\ H\&K,  \SII\,, \SiII\, and \CaII\, IRT. 
The profiles and velocities of \SiII\, \ld6355 absorption features are approximately the same for all samples.
The pseudo-equivalent widths (pEWs) of \SiII\, \ld5972, 6355 measured at this phase for SN~2024gy are 15\,\AA\ and  103\,\AA, suggesting that it can be put into the core-normal (CN) subgroup within Branch’s classification scheme \citep{2006PASP..118..560B}.

At around one month after maximum light,  the spectra of SN 2024gy and other samples are very similar, dominated by the characteristic lines of \FeII.

\subsection{Late-Time Spectra}\label{subsec:late-time-spec}
\begin{figure*}
    \centering
    \includegraphics[width=0.95\textwidth]{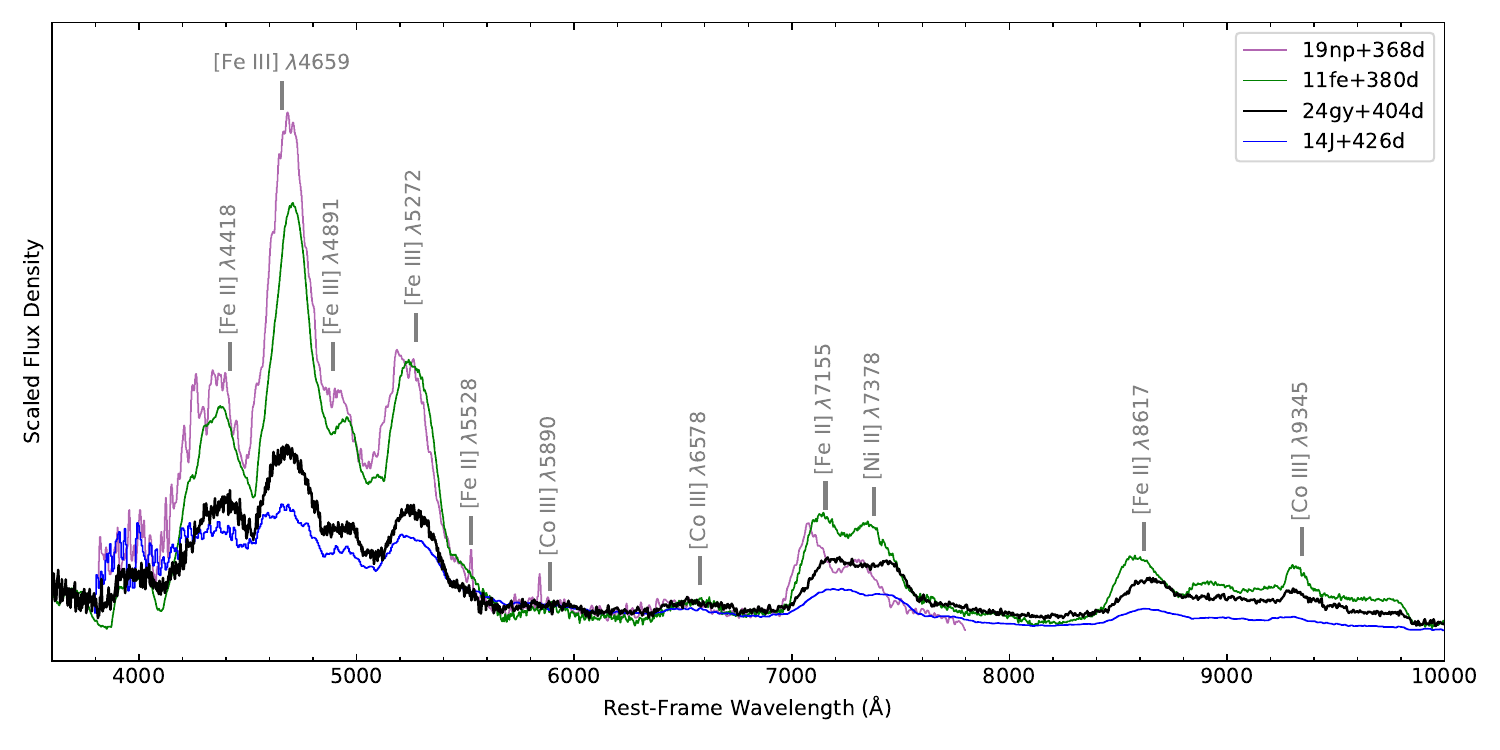}
    \caption{Late-time spectra at $+404$ d of SN 2024gy (black). Nebular spectra of SN 2011fe, SN 2014J, and SN 2019np at similar phases are shown for comparison. All spectra have been corrected for host-galaxy redshifts and reddening. Important emission lines are marked and labeled.
    }
    \label{fig:latespec}
\end{figure*}

In the nebular phase of SN 2024gy ($>+300$d), the spectra are dominated by strong emission features, with all absorption lines (e.g., \CaII\,IRT) having given way to emission (Figure \ref{fig:spec}). 
For example, the optical spectrum observed at $+404$ d after \textit{B}-band maximum is displayed in Figure~\ref{fig:latespec}.
For comparison, the late-time spectra of SN 2019np ($+368$d), SN 2011fe ($+380$d), and SN 2014J ($+426$d) are also plotted.
All spectra are reddening-calibrated and scaled to the flux density minimum.
The features in the spectrum of SN 2024gy at $+404$ d are similar to those of other SNe~Ia, manifesting the dominance of iron-group  forbidden emission lines.
The relative fluxes of the Fe emission features, predominantly spanning 4200-5600\,\AA, exhibit a monotonic temporal decay.
This may be caused by the transition phase of the nebula, cooling from optical lines to  mid-infrared lines \citep[][]{2015ApJ...814L...2F}.
In addition, the flux ratios of [\FeIII] \ld4659 to [\FeII] \ld5272 seem to become progressively smaller.
This behavior indicates that $\sim 400$ days post maximum light may be the phase of iron-group ionization state transitioning (from doubly to singly ionized) for SNe Ia.

The [\FeII]/[\NiII]-dominated region around 7300\,\AA\, could be used to infer the iron and Ni abundances.
With some reasonable assumptions, one may put constraints on the explosion mechanism with this method \citep[e.g.,][]{2018MNRAS.477.3567M,2023MNRAS.526.1268L}. 
We carefully perform such an analysis on the spectrum of SN 2024gy at $+404$ d, considering the better resolution.
There are four [\FeII] lines (\ld7155, 7172, 7388, 7453) and two [\NiII] lines (\ld7378, 7412) concentrated in this region.
After continuum subtraction, we fit a six-component (corresponding to six lines) Gaussian function  to the spectrum around 7300\,\AA, following the parameter setting of \citet{2023MNRAS.526.1268L}.
In short, the amplitudes of [\FeII] and [\NiII] lines are fixed to specific proportions to that of [\FeII] \ld7155 and [\NiII] \ld7378, respectively.
The line widths of the same element (e.g., [\FeII] group) are set to be identical.
There are thus only six free parameters in the fitting.

\begin{figure}
    \centering
    \includegraphics[width=\columnwidth]{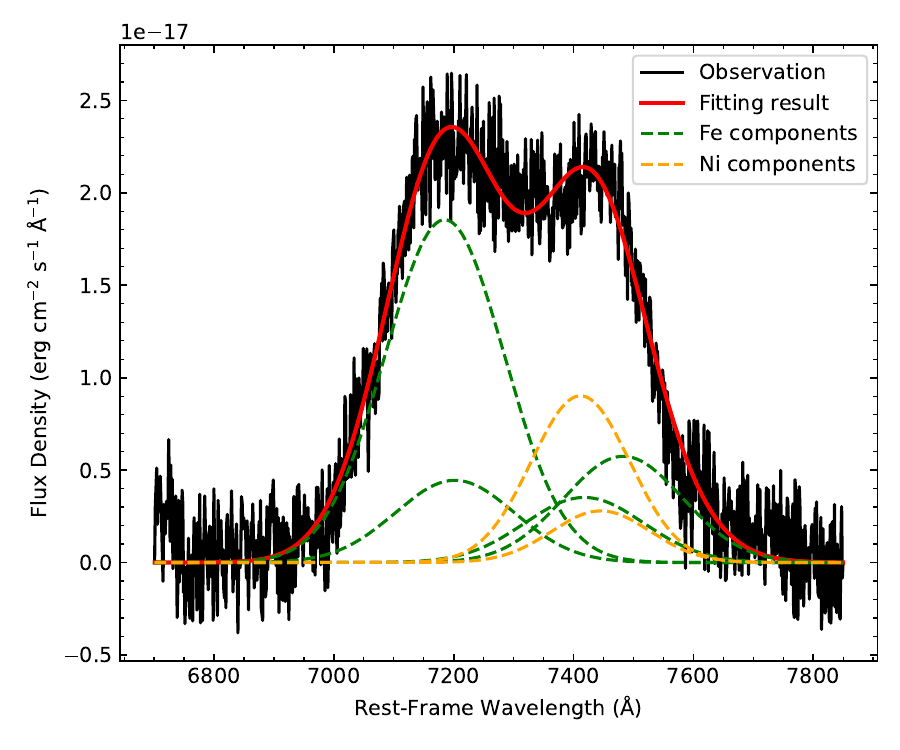}
    \caption{Fitting results of the late-time spectra (+404 d) of SN~2024gy around 7300\,\AA. Reddening-corrected spectra are shown in black. The overall fits are shown with a red line, while the [\FeII] and [\NiII] features are represented by green and yellow dashed lines, respectively. The continuum of the observed spectrum is reduced using a simple linear function.
    }
    \label{fig:Ni_Fe}
\end{figure}

The fitting results are displayed in Figure \ref{fig:Ni_Fe}.
According to the optimized parameters, we find that the velocity shifts of [\FeII] \ld7155 and [\NiII] \ld7378 are $1255\pm130~\rm km~s^{-1}$ and $1370\pm150~\rm km~s^{-1}$, respectively.
The full width at half-maximum intensity (FWHM)  of [\FeII] \ld7155 and [\NiII] \ld7378 are $9680\pm220~\rm km~s^{-1}$ and $7770\pm520~\rm km~s^{-1}$.
The derived values are consistent with the average values obtained from the late-time SN~Ia samples reported by \citet{2018MNRAS.477.3567M} or \citet{2023MNRAS.526.1268L}, demonstrating that our fitting captures the [\FeII] and [\NiII] emission appropriately for SN 2024gy.
Using the flux ratio of [\NiII] \ld7378 and [\FeII] \ld7155, one can estimate the Ni/Fe abundance ratio with reasonable parameters of local thermodynamic equilibrium in the corresponding temperature at this phase.
Following the calculation of \citet{2018MNRAS.477.3567M}, we derive the Ni/Fe mass ratio of $M_{\rm Ni}/M_{\rm Fe}=0.080_{-0.036}^{+0.053}$ for SN 2024gy at +404 d.
The uncertainty is derived from 10,000 Monte Carlo simulations at the 95\% confidence level by varying the parameters described in \cite{2018MNRAS.477.3567M}. 
This value probably falls within the Ni/Fe ratio range \citep[$\sim 0.06$--0.11;][]{2023MNRAS.526.1268L} predicted by the delayed-detonation model \citep[][]{2013MNRAS.429.1156S}.
A consistent ratio is derived from the +320 d spectrum.

\subsection{The Evolution of \NaI\,} \label{subsec:NaID}

\begin{figure}
    \centering
    \includegraphics[width=\columnwidth]{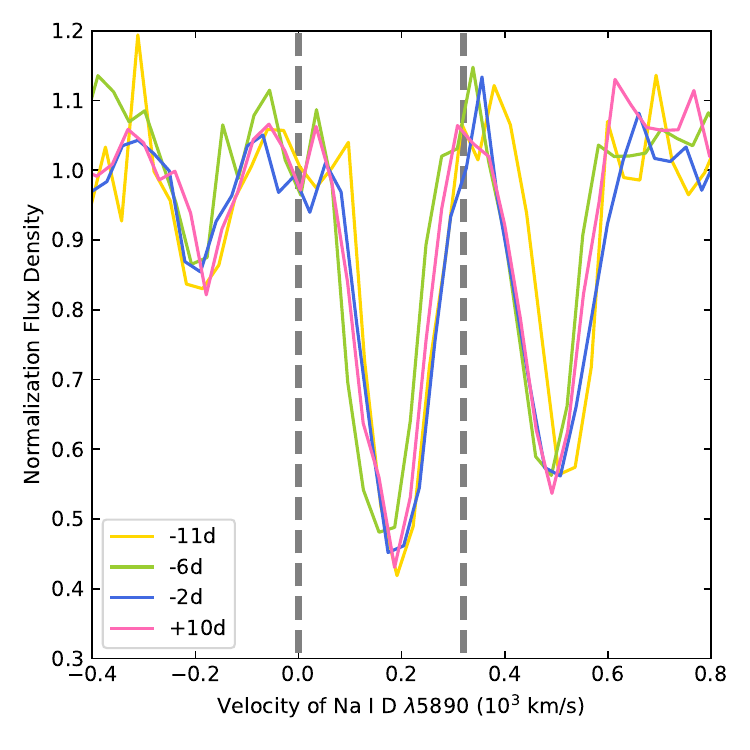}
    \caption{The evolution of the \NaI\, doublet lines in four cross-dispersion spectra, with different colors representing different periods. The abscissa displays the velocity of \NaI\, \ld5890.
    }
    \label{fig:Na}
\end{figure}

We measure the EWs of \NaI\, in all spectra and investigate whether there is an evolutionary trend.
However, the HVF of \SiII\, \ld6355 and \NaI\, are mixed together in the early low-resolution spectra, making it difficult to measure  accurate EWs of the \NaI.
Furthermore, the resulting EW values using spectra observed at similar times with different telescopes show significant discrepancies.

We plot the evolution of the \NaI\, doublet in the mid-resolution spectra in Figure~\ref{fig:Na}. The phases of these four spectra include before and after maximum light.
It seems that the EW of \NaI\, does not significantly exhibit temporal evolution.
We also measure the velocity of the \NaI\, finding no blueshift or redshift.
The invariable evolution behaviors indicate that the \NaI\, absorption might caused by the interstellar medium (ISM) rather than the CSM.
Indeed, the \NaI\, shown in the spectrum of the host galaxy (NGC 4216) is comparable to that of SN 2024gy (see Section \ref{subsec:reddening}), also suggesting the ISM origin for the  \NaI.

\subsection{Ejecta Velocity}
\begin{figure}
    \centering
    \includegraphics[width=\columnwidth]{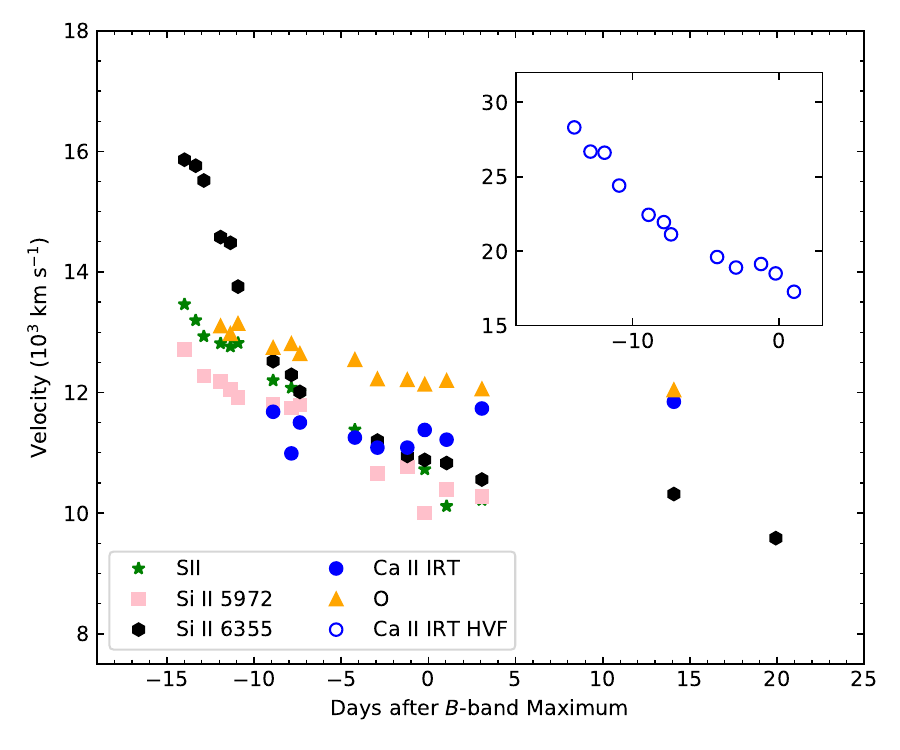}
    \caption{Evolution of the expansion velocity of SN~2024gy as measured from the absorption minimum of \SiII\, \ld5972, \SiII\, \ld6355, \SII\, \ld5468, \OI\,\ld7773, and \CaII\, IRT. 
    }
    \label{fig:vel}
\end{figure}

We measure the ejecta velocity of SN 2024gy through the absorption minimum of selected lines, including \SiII\, \ld6355, \SII\, \ld5468, \OI\,\ld7773, and \CaII\, IRT. The velocity evolution is shown in Figure \ref{fig:vel}. We obtain those velocities by using Gaussian fitting to measure the wavelength of these blueshifted spectral lines at the point of minimum absorption.

According to the velocity of \SiII\, \ld6355 at maximum light, measured as 11,200 km $\rm s^{-1}$, SN 2024gy can be placed into the normal-velocity (NV) group classification scheme proposed by \citet{2009ApJ...699L.139W}. The velocity gradient of \SiII\, \ld6355 is found to be $31 \pm 12$ km $\rm s^{-2}$ at +10 days, indicating that SN 2024gy belongs to the low velocity gradient (LVG) subgroup \citep{2005ApJ...623.1011B}. 

\begin{figure*}
    \centering
    \includegraphics[width=0.49\linewidth]{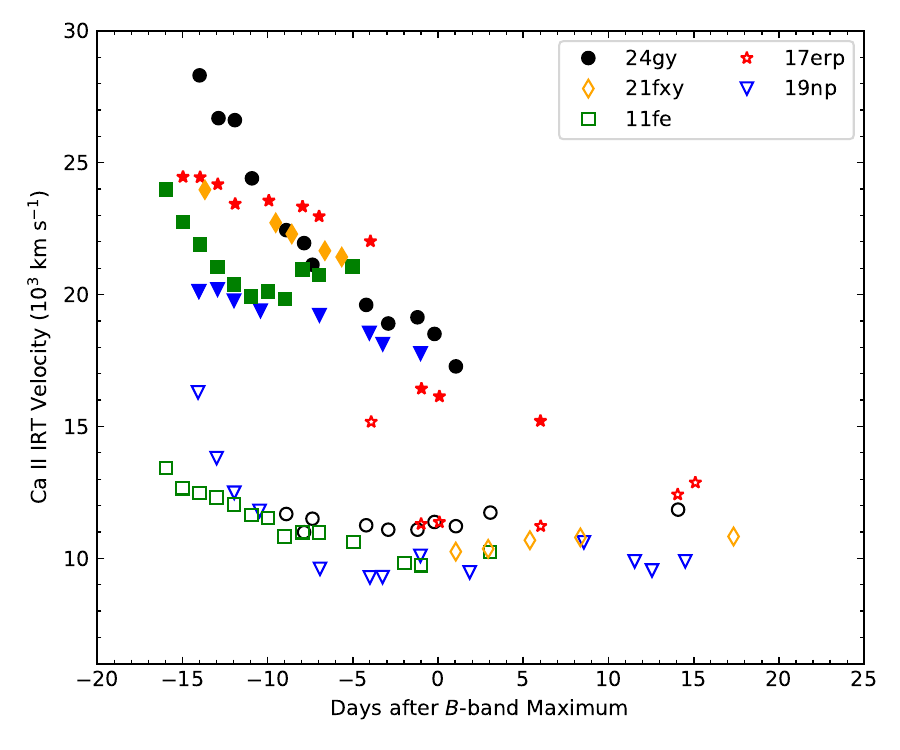}
    \includegraphics[width=0.49\linewidth]{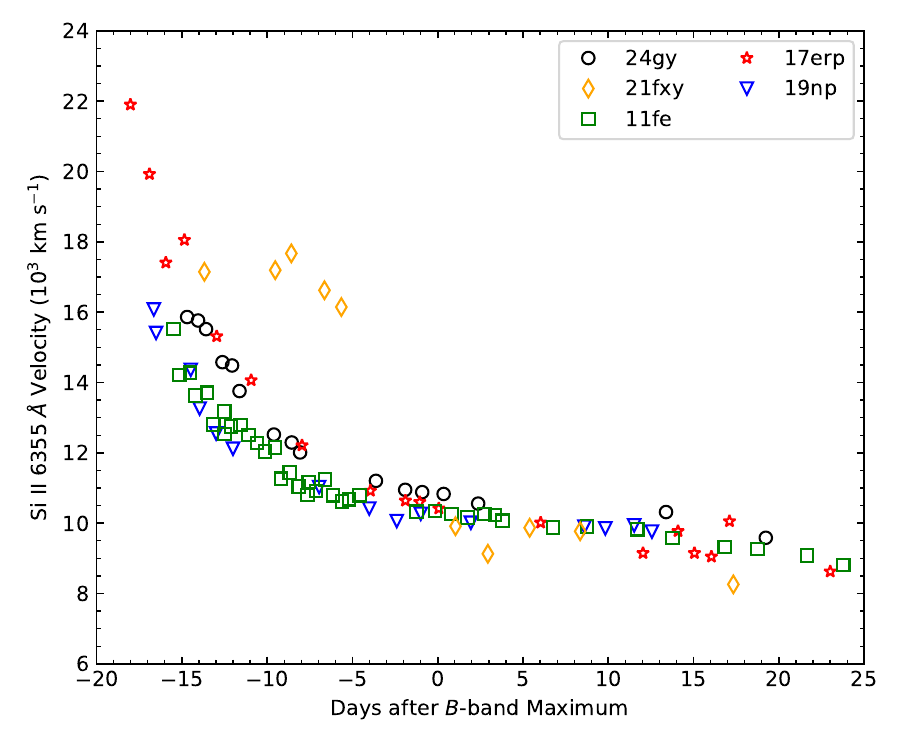}
    \caption{Velocity comparison of \SiII\, \ld6355 and \CaII\, IRT among SN 2024gy (black) and with other well-sampled SN Ia. Solid markers represent HVFs, while hollow markers represent PV features.}
    \label{fig:vel_com}
\end{figure*}

We also plot the velocity comparison of \SiII\, \ld6355 and \CaII\, IRT among each sample, as shown in Figure~\ref{fig:vel_com}.
One can see that the $v_{\rm Si}$ and $v_{\rm Ca}$ evolution (including that of the HVFs) of SN 2024gy is similar to that of other SNe Ia.
This indicates that SN 2024gy may share a comparable explosion energy and expansion  (e.g., homologous expansion). 

\section{discussion} \label{sec:5}
\subsection{Absolute Magnitude and Bolometric Luminosity}

The host galaxy NGC 4216 exhibits a significant dispersion in its distance-modulus values within the NED, ranging from $\mu \approx$\,30.65 to 31.46 mag. This $\sim$\,0.8 mag spread (corresponding to a distance uncertainty of $\sim$\,3.5 Mpc) primarily stems from heterogeneous measurement techniques in historical datasets. We choose to select the most recent Tully-Fisher measurement of $\mu = 30.89 \pm 0.12$ mag from \citet{2020ApJ...902..145K} as the NED result for further calculation.

Through SALT2 light-curve fitting combined with the EBV\_model in SNooPy, we derive a distance modulus of $\mu=31.57\pm0.16$ mag for SN 2024gy.
This value is slightly larger than that of the Tully-Fisher measurement.
For subsequent analysis, we adopt the averaged value of $\mu\,= 31.23 \pm\,0.14$ mag as the distance modulus for SN 2024gy.

Using $E(B-V)_{\rm host}=0.38\pm0.10$ mag and $R_V^{\rm host}=1.5$ as the host reddening parameters, the $B$-band host extinction $A_B^{\rm host}$ is calculated to be $0.95 \pm 0.25$ mag.
The total $A_B$ is then $1.07 \pm 0.25$ mag, with a Galactic extinction of $A_B^{\rm Gal}=0.115$ mag. Therefore, the absolute \textit{B}-band maximum is $-19.25\pm0.29$ mag, combining the parameters adopted above.

To estimate the peak bolometric luminosity, we construct the spectral energy distribution using \textit{uBVri}-band photometry for SN 2024gy from LJT. 
Following standard bolometric corrections for SNe Ia, we account for unobserved spectral regions by adding 15\% of the optical flux for the UV contribution \citep{2012ApJ...749..126W} and 5\% for the near-infrared component \citep{2009ApJ...697..380W}. This yields a peak bolometric luminosity of $(1.2\pm0.3)\times10^{43}\ \rm{erg\ s^{-1}}$ for SN 2024gy.

We estimate the synthesized $^{56}$Ni mass through the ``Arnett law" \citep{1982ApJ...253..785A,2005A&A...431..423S}, combining our derived luminosity with the light-curve rise time. The rise time is determined by fitting a $t^2$ model to the quasibolometric light curve (constructed from $uBVri$ photometry) during the premaximum phase ($t\leq -10$ d), following the methodology of \citet{2012ApJ...745...44G} and \citet{2015MNRAS.446.3895F}. The best-fit rise time of  $t_{\rm r} = 17.4 \pm 0.3$ d, when combined with our bolometric luminosity measurement, gives a $^{56}$Ni mass estimate of $ 0.57\pm0.14\ {\rm M}_{\sun}$ for SN 2024gy.

The derived parameters for SN 2024gy, including its peak absolute magnitude, bolometric luminosity, rise time, and nickel mass, firmly establish its classification as a normal SN Ia. These values show remarkable consistency with the well-studied prototypical SN 2011fe \citep[$M_{\rm max}(B)\approx -19.24$ mag, $L_{\rm peak}\approx 1.13\times10^{43}\ \rm{erg\ s^{-1}}$, $t_{\rm r} \approx 17.64$ days, $M_{\rm ^{56}Ni} \approx 0.57\ M_{\sun}$;][]{2016ApJ...820...67Z}, as further evidenced by the nearly identical $B$ and $V$ light-curve evolution shown in Figure~\ref{fig:LCcom}.

However, detailed spectroscopic analysis (Section~\ref{sec:4}) reveals significant deviations in the early-phase behavior, particularly regarding (1) enhanced IME absorption features, and (2) exceptionally high-velocity components in the \CaII\, IRT. These spectroscopic peculiarities suggest that, while SN 2024gy follows the standard photometric template for normal SNe Ia, its explosion dynamics or progenitor system may have distinctive characteristics that warrant further investigation.

Our analysis reveals that the observed diversity in HVFs within the outer ejecta layers does not significantly influence the peak luminosity of SNe~Ia. While these HVF variations reflect fundamental differences in ejecta properties that should correspond to distinct explosion mechanisms, their impact on the peak luminosity appears negligible. This is particularly evident as HVF signatures typically become undetectable near maximum light, suggesting that the inner ejecta layers visible during this epoch exhibit remarkable homogeneity across normal SNe Ia.

This observed uniformity has two important implications. On the one hand, it provides the physical basis for the successful use of SNe Ia as standardizable candles in cosmological distance measurements. On the other hand, it indicates that studies aiming to probe explosion mechanisms through ejecta structure must focus on early-phase (e.g., $t \leq\,-5$ d) spectra, before expansion and mixing processes homogenize the ejecta properties.


\subsection{The Origin of the High-Velocity Features}

The early-phase spectra of SN 2024gy exhibit detached HVFs in the \CaII\,IRT that appear remarkably deeper and broader than those observed in comparison SNe Ia (Figure~\ref{fig:spcom}). At $t \approx -15$ d, this feature shows exceptionally high velocities exceeding $25,000\, \rm km\ s^{-1}$, ranking among the most extreme HVFs ever recorded for SNe Ia. Similar high-velocity components have been documented in several well-studied SNe Ia, including SN 2009ig \citep{2013ApJ...777...40M}, SN 2012fr \citep{2014AJ....148....1Z}, SN 2017erp \citep{2019ApJ...877..152B}, and SN 2019ein \citep{2020ApJ...897..159P,2022MNRAS.517.4098X}.

Current theoretical frameworks attribute HVF formation to physical processes occurring in the outermost ejecta layers, with three primary mechanisms under consideration: abundance enhancements (AE), density enhancements (DE), and ionization effects (IE) \citep[e.g.,][]{2005MNRAS.357..200M}. 

The delayed-detonation model provides a natural explanation for potential AE in the high-velocity layers. In this scenario, the explosion could produce high-velocity ejecta ($\sim$\,18,000--30,000 km s$^{-1}$) during early phases through slight modifications to the outermost material distribution \citep{2014MNRAS.439.1959M,2020ApJ...893..143K}.  For SN 2024gy, the \CaII\,IRT features might result from AE in the outermost layers via possible outward mixing, a mechanism invoked to explain HVFs in objects like SN 2019ein \citep{2020ApJ...897..159P}. However, the exceptionally broad and deep \CaII\, absorption in SN 2024gy suggests either a more complex density structure in the outermost layers or a special stellar environment.


The double-detonation scenario offers an alternative explanation, where detonation in a helium shell produces AE and creates an intrinsically asymmetric explosion that generates high-velocity effects at certain viewing angles. While this scenario can explain early flux excesses observed in some SNe Ia \citep[e.g., SN 2015bq,][]{2022ApJ...924...35L} through mechanisms described by \citet{2018ApJ...865..149J} and others, no such early excess was detected in SN 2024gy. Recent studies show that reducing the helium shell mass to $\lesssim 0.02 ~\rm M_{\sun}$ in double-detonation models can reproduce normal SN Ia light curves without early excesses \citep[][]{2019ApJ...878L..38T,2021ApJ...919..126B}.
For $M_{\rm ^{56}Ni} \approx 0.6\, \rm M_{\sun}$, this configuration yields an IME mass of $\sim 0.3\, \rm M_{\sun}$,  significantly less than the ($\sim 0.5\, \rm M_{\sun}$) produced in delayed-detonation models \citep[e.g.,][]{2013MNRAS.429.1156S,2022A&A...659A..27L}. 

The early-time spectra of SN 2024gy exhibit deeper IME absorption features, particularly in the \SiII\, lines, indicating enhanced photospheric IME production. This signature favors a delayed-detonation origin, likely resulting from an extended deflagration phase that facilitates more complete nuclear burning. 
Furthermore, the late-time spectra of SN 2024gy could be consistent with the delayed-detonation scenario, as indicated by the derived Ni/Fe ratio (Section \ref{subsec:late-time-spec}).
In addition, a study on the infrared spectra at the nebular phase, which appeared during the review process of this paper, also suggests a delayed-detonation explosion for SN 2024gy \citep{2025arXiv251009760K}.

Independent of progenitor-related AE, the outermost explosion zone could demonstrate significant HVFs through combined DE and IE effects.
The interaction between normal outermost ejecta and relatively dense CSM, such as clumpy clouds or a shell \citep[e.g.,][]{2004ApJ...607..391G,2017MNRAS.467..778M}, result in DE and IE that cause the broad and deep HVFs \citep[][]{2005MNRAS.357..200M}.
Observational evidence for CSM includes blueshifted \NaI\, lines in some SNe Ia \citep[e.g.,][]{2011Sci...333..856S,2013MNRAS.436..222M,2019ApJ...882..120W}, with temporal variations in \NaI\, features providing strong evidence for the presence of CSM \citep[][]{2016A&A...592A..40F,2019ApJ...882..120W}. However, SN 2024gy shows strong but stable \ion{Na}{1} D absorption without significant velocity or temporal variations, suggesting an ISM origin as discussed in Section \ref{subsec:NaID}.

If broad HVFs require IE in the high-velocity zone \citep[][]{2005MNRAS.357..200M}, the presence of low-mass ($\sim10^{-3}\, \rm M_{\sun}$) hydrogen-rich CSM becomes plausible \citep[][]{2019ApJ...886...58M}. 
Assuming there is CSM of $\sim10^{-3}\, \rm M_{\sun}$ from the SN to a radius of $\sim 0.1$\,pc, the column density of Na in the CSM can be estimated as $\sim 8\times10^{11} ~\rm{cm^{-2}}$ according to the shell model with other ordinary parameters given by \citet{2009ApJ...699L..64B}.
Such a column density could cause \NaI\, absorption with EW $\approx 0.5$\,\AA\  \citep[e.g.,][]{2013ApJ...779...38P}. 
If the \NaI\, absorption is significantly ($> 10\%$) affected by the shock-CSM interaction, the EW variation is $\sim0.1$\,\AA.
This variation is negligible and undetectable for SN 2024gy, since the EW of \ion{Na}{1}~D that is mostly attribute to the ISM is $\sim 3$~\AA\ in our low-resolution spectra.
Thus, the scenario that contains low-mass CSM is reasonable for SN 2024gy.

In this model, \CaIII\, initially present in high-velocity zones could recombine to form \CaII\, after the explosion, with hydrogen providing sufficient free electrons in the ejecta.
This process might naturally explain the broad, deep \CaII\,HVFs in SN 2024gy.
In addition, the HVFs of \SiII, which are absent in SN 2024gy, could also be involved in this scenario.
The possible \SiIII\, in the high-velocity zone seems not to recombine to form \SiII\, at the same temperature, according to the different excitation energy \cite[e.g.,][]{2015ApJS..220...20Z}.

In summary, these findings suggest that the extreme \CaII\, HVFs observed in SN 2024gy are likely produced by delayed detonation in the outermost ejecta layers, where the hydrogen of CSM is potentially mixed to suppress the higher ionization.  Future investigations should focus on further quantifying the physics of recombination and ejecta stratification to refine our understanding of these processes in SNe Ia.

\section{conclusion} \label{sec:6}
We present photometric and spectroscopic observations of SN 2024gy in NGC 4216 ($z=0.00043$). The overall spectral evolution resembles that of normal SNe~Ia, while prominent \NaI\, absorption is attributed to the ISM rather than the CSM given the absence of temporal evolution. Multiple methods establish substantial host-galaxy reddening with $E(B-V)=0.38\pm0.10$ mag and a low extinction ratio of $R_V\approx 1.5$.

Adopting the derived distance modulus of $\mu= 31.23\pm0.26$ mag, we calculate an absolute $B$-band peak brightness of $-19.25\pm0.29$ mag for SN 2024gy. This luminosity is fully consistent with the Phillips relation given its light-curve decline rate, $\Delta m_{15}(B) = 1.12\pm0.04$ mag. We estimate a peak bolometric luminosity of $L_{\rm max} = (1.2\pm0.3)\times10^{43}\ \rm{erg\ s^{-1}}$, corresponding to a synthesized $0.57\pm0.14\ {\rm M}_{\sun}$ of $^{56}$Ni. 

The multi-epoch spectroscopy of SN 2024gy reveals possible evidence for a delayed-detonation mechanism. 
The early-phase spectra exhibit extraordinarily strong HVFs in the \CaII\, IRT, with velocities exceeding 25,000 $~\rm km~s^{-1}$, contrasting with the significantly lower \SiII\,\ld\,6355 velocity of $\sim 16,000 ~\rm km~s^{-1} $ at the same epoch. This velocity disparity, coupled with deeper \SiII\,absorption indicative of enhanced IME synthesis, points to deflagration burning. The absence of \SiII\, HVFs alongside prominent \CaII\,HVFs implies distinct ionization states in the outermost ejecta, likely due to minimal hydrogen mixing suppressing higher ionization --- a signature consistent with delayed detonation.
Furthermore, the Ni/Fe mass ratio estimated from the nebular spectroscopy might aligns with the range predicted by delayed-detonation models.
These early and late spectroscopic diagnostics form a possible evidence chain on delayed detonation as the explosion mechanism for SN 2024gy.

In summary, the photometric properties of SN 2024gy are similar to that of normal SNe Ia.
The spectroscopic evolution from early high-velocity features to late-time nucleosynthetic products indicate a possible delayed detonation.

Future high-cadence spectroscopic campaigns should quantify the physics of HVF formation and constrain progenitor configurations through similar multi-epoch diagnostics.

\section*{Acknowledgments}
This work is supported by the National Key R\&D Program of China with grant 2021YFA1600404, the National Natural Science Foundation of China (NSFC grants 12173082, 12333008, and 12225304), the Yunnan Fundamental Research Projects (YFRP; grants 202501AV070012, 202401BC070007, 202201AT070069, and 202501AS070005), the Top-notch Young Talents Program of Yunnan Province, the Light of West China Program provided by the Chinese Academy of Sciences (CAS), and the International Centre of Supernovae, Yunnan Key Laboratory (grant 202302AN360001). 
X.F. Wang is supported by the NSFC (grants 12288102, 12033003, and 11633002) and the Tencent Xplorer Prize. 
A.V.F.’s research group at U.C. Berkeley acknowledges financial                 
assistance from the Christopher R. Redlich Fund, as well as donations         
from Gary and Cynthia Bengier, Clark and Sharon Winslow, Alan Eustace and Kathy Kwan,        
William Draper, Timothy and Melissa Draper, Briggs and Kathleen Wood, Sanford Robertson (W.Z. is a Bengier-Winslow-Eustace Specialist
in Astronomy, T.G.B. is a Draper-Wood-Robertson Specialist in                
Astronomy), and numerous other donors. 
L.-Z. Wang is sponsored by the National Natural Science Foundation of China (NSFC) grants No. 12573050,  the Chinese Academy of Sciences South America Center for Astronomy (CASSACA) Key Research Project E52H540301, and  in part by the Chinese Academy of Sciences (CAS) through a grant to the CASSACA.
B.K. acknowledges support from the ``Special Project for High-End Foreign Experts," Xingdian Funding from Yunnan Province, and the Key Laboratory of Survey Science of Yunnan Province (grant 202449CE340002). L.-Z. Wang is supported by the CAS South America Center for Astronomy (CASSACA) Key Research Project E52H540301, and in part by CAS through a grant to the CASSACA.
Y.-Z. Cai is supported by the NSFC (grant 12303054) and the YFRP (grants 202401AU070063 and 202501AS070078).
B.K. and Y.-Z. Cai also acknowledge the National Key Research and Development Program of China (grant 2024YFA1611603).
A.R. acknowledges financial support from the GRAWITA Large Program Grant (PI P. D'Avanzo) and from the PRIN-INAF 2022 ``Shedding light on the nature of gap transients: from the observations to the models.''
O.C. acknowledges Dr. R Michael Rich and the McNair Research Scholars Program for their invaluable mentorship and support.
We acknowledge Carl Melis and Tony Rodriguez for providing the spectra obtained by Shane telescope on Jan 6 and Jan 8, 2024 respectively.

We acknowledge the support of the staff of the LJT, XLT, Keck Observatory, and Lick Shane telescope.
Funding for the LJT has been provided by the CAS and the People's Government of Yunnan Province. The LJT is jointly operated and administrated by YNAO and Center for Astronomical Mega-Science, CAS. 
Mephisto is developed at and operated by the South-Western Institute for Astronomy Research of Yunnan University (SWIFAR-YNU), funded by the ``Yunnan University Development Plan for World-Class University" and ``Yunnan University Development Plan for World-Class Astronomy Discipline."
Partly based on observations collected at the Schmidt 67/92\,cm telescope operated by INAF -- Osservatorio Astronomico di Padova at Cima Ekar, Asiago, Italy.
Some of the data presented herein were obtained at the W. M. Keck
Observatory, which is operated as a scientific partnership among the
California Institute of Technology, the University of California, and
NASA; the observatory was made possible by the generous financial
support of the W. M. Keck Foundation.
A major upgrade of the Kast spectrograph on the Shane 3\,m telescope at     
 Lick Observatory, led by Brad Holden, was made possible through               
generous gifts from the Heising-Simons Foundation, William and Marina
Kast, and the University of California Observatories.  Research at        
Lick Observatory is partially supported by a generous gift from Google.

We thank the anonymous referee for the constructive comments that helped to improve the presentation of the paper.
%

\bibliography{24gy}{}
\bibliographystyle{aasjournal}

\appendix
\restartappendixnumbering

\section{Photometric and Spectroscopic Data}
The following are the photometric and spectroscopic data for SN 2024gy, including the Johnson \textit{BV}, Sloan \textit{ugri}, and near-infrared \textit{H} photometry (Table \ref{app_a:photo}), Mephisto photometry (Table \ref{app_a:M_photo}), and the journal of spectral observations (Table \ref{app_a:spec}). Note that the Mephisto data observed in the same day are binned into a single data point. Data with the original temporal resolution are shown in Figure \ref{fig:LC1}. The effective wavelength coverages of the Mephisto \textit{uvgriz} bands are described in detail  in Section \ref{subsec:photometry}.

\startlongtable
\begin{deluxetable*}{lcccccccccc}
\tablecaption{Johnson and Sloan Photometry of SN 2024gy\label{app_a:photo}}
\tabletypesize{\scriptsize}
\tablewidth{0pt}
\tablehead{
\colhead{MJD} & \colhead{Phase$^a$} & \colhead{\textit{u}} & \colhead{\textit{B}} & \colhead{\textit{V}}& \colhead{\textit{g}}& \colhead{\textit{r}}& \colhead{\textit{i}}& \colhead{\textit{z}}& \colhead{\textit{H}}& \colhead{Telescope}\\
\colhead{} & \colhead{days} & \colhead{mag} & \colhead{mag} & \colhead{mag}& \colhead{mag}& \colhead{mag}& \colhead{mag}& \colhead{mag}& \colhead{mag}& \colhead{}
}
\startdata
60314.84 	&	-14.7 	&	17.64(04)	&	16.22(08)	&	15.53(08)	&	15.86(09)	&	15.44(06)	&	15.94(05)	&	15.28(08)	&	$\cdots$	&	LJT	\\
60314.87 	&	-14.6 	&	$\cdots$	&	16.33(10)	&	15.52(07)	&	15.88(05)	&	15.46(04)	&	15.92(06)	&	$\cdots$	&	$\cdots$	&	TNT	\\
60314.94 	&	-14.6 	&	$\cdots$	&	16.17(08)	&	15.48(07)	&	15.81(08)	&	15.44(07)	&	15.96(07)	&	15.15(01)	&	$\cdots$	&	LJT	\\
60315.24 	&	-14.3 	&	$\cdots$	&	$\cdots$	&	$\cdots$	&	15.59(02)	&	15.18(02)	&	15.64(02)	&	14.87(02)	&	13.96(09)	&	REM	\\
60315.88 	&	-13.6 	&	$\cdots$	&	15.60(22)	&	15.03(13)	&	15.29(18)	&	14.93(09)	&	15.45(14)	&	$\cdots$	&	$\cdots$	&	TNT	\\
60315.94 	&	-13.6 	&	16.86(02)	&	15.42(08)	&	14.91(05)	&	15.15(05)	&	14.85(06)	&	15.35(05)	&	14.70(02)	&	$\cdots$	&	LJT	\\
60316.28 	&	-13.2 	&	$\cdots$	&	$\cdots$	&	$\cdots$	&	15.01(02)	&	14.69(02)	&	15.02(02)	&	14.55(02)	&	$\cdots$	&	REM	\\
60316.80 	&	-12.7 	&	$\cdots$	&	15.07(21)	&	14.64(12)	&	14.82(17)	&	14.50(07)	&	14.97(12)	&	$\cdots$	&	$\cdots$	&	TNT	\\
60316.90 	&	-12.6 	&	16.15(02)	&	14.93(09)	&	14.55(12)	&	14.70(06)	&	14.51(07)	&	14.97(08)	&	14.21(01)	&	$\cdots$	&	LJT	\\
60317.90 	&	-11.6 	&	15.57(01)	&	14.52(08)	&	14.22(09)	&	14.34(09)	&	14.13(07)	&	14.52(07)	&	14.06(01)	&	$\cdots$	&	LJT	\\
60318.25 	&	-11.3 	&	$\cdots$	&	$\cdots$	&	$\cdots$	&	14.35(02)	&	14.09(02)	&	14.29(02)	&	13.87(02)	&	$\cdots$	&	REM	\\
60318.87 	&	-10.6 	&	$\cdots$	&	14.25(21)	&	13.96(11)	&	14.08(17)	&	13.80(07)	&	14.15(13)	&	$\cdots$	&	$\cdots$	&	TNT	\\
60318.95 	&	-10.5 	&	15.08(01)	&	$\cdots$	&	$\cdots$	&	$\cdots$	&	$\cdots$	&	$\cdots$	&	13.78(10)	&	$\cdots$	&	LJT	\\
60319.25 	&	-10.3 	&	$\cdots$	&	$\cdots$	&	$\cdots$	&	13.97(02)	&	13.72(02)	&	14.02(02)	&	13.70(02)	&	12.92(04)	&	REM	\\
60319.83 	&	-9.7 	&	$\cdots$	&	13.97(23)	&	13.73(12)	&	13.84(17)	&	13.57(07)	&	13.87(13)	&	$\cdots$	&	$\cdots$	&	TNT	\\
60319.91 	&	-9.6 	&	14.75(01)	&	13.86(07)	&	13.68(09)	&	13.79(11)	&	13.58(06)	&	13.90(10)	&	13.64(01)	&	$\cdots$	&	LJT	\\
60320.28 	&	-9.2 	&	$\cdots$	&	$\cdots$	&	$\cdots$	&	14.01(02)	&	13.68(02)	&	13.74(02)	&	13.68(02)	&	$\cdots$	&	REM	\\
60320.88 	&	-8.6 	&	$\cdots$	&	13.81(20)	&	13.54(12)	&	13.65(16)	&	13.39(07)	&	13.67(12)	&	$\cdots$	&	$\cdots$	&	TNT	\\
60320.96 	&	-8.5 	&	14.44(01)	&	$\cdots$	&	$\cdots$	&	13.55(09)	&	13.39(07)	&	13.72(08)	&	$\cdots$	&	$\cdots$	&	LJT	\\
60321.36 	&	-8.1 	&	$\cdots$	&	$\cdots$	&	$\cdots$	&	13.54(02)	&	13.43(02)	&	13.44(02)	&	13.32(02)	&	12.50(02)	&	REM	\\
60321.89 	&	-7.6 	&	$\cdots$	&	13.65(23)	&	13.40(16)	&	13.48(16)	&	13.25(05)	&	13.53(11)	&	$\cdots$	&	$\cdots$	&	TNT	\\
60322.48 	&	-7.0 	&	$\cdots$	&	$\cdots$	&	$\cdots$	&	13.28(02)	&	13.11(03)	&	$\cdots$	&	$\cdots$	&	$\cdots$	&	ZTF	\\
60322.94 	&	-6.6 	&	14.06(01)	&	13.35(07)	&	13.18(07)	&	13.24(07)	&	13.13(06)	&	13.43(12)	&	13.30(01)	&	$\cdots$	&	LJT	\\
60323.00 	&	-6.5 	&	14.08(03)	&	13.24(03)	&	13.09(03)	&	13.20(03)	&	13.00(04)	&	13.27(04)	&	$\cdots$	&	$\cdots$	&	Schmidt	\\
60323.23 	&	-6.3 	&	$\cdots$	&	$\cdots$	&	$\cdots$	&	13.34(02)	&	13.10(02)	&	13.25(02)	&	13.28(02)	&	12.66(03)	&	REM	\\
60323.80 	&	-5.7 	&	$\cdots$	&	13.29(11)	&	13.12(09)	&	13.18(08)	&	13.03(07)	&	13.30(08)	&	$\cdots$	&	$\cdots$	&	TNT	\\
60323.87 	&	-5.6 	&	14.01(01)	&	13.26(09)	&	13.06(07)	&	13.14(08)	&	13.01(05)	&	13.34(06)	&	13.27(01)	&	$\cdots$	&	LJT	\\
60324.79 	&	-4.7 	&	$\cdots$	&	13.24(12)	&	13.03(10)	&	13.09(10)	&	12.96(09)	&	13.27(09)	&	$\cdots$	&	$\cdots$	&	TNT	\\
60324.93 	&	-4.6 	&	13.85(01)	&	13.15(05)	&	12.97(11)	&	13.05(07)	&	12.91(13)	&	13.31(09)	&	13.22(01)	&	$\cdots$	&	LJT	\\
60325.26 	&	-4.2 	&	$\cdots$	&	$\cdots$	&	$\cdots$	&	13.33(02)	&	13.07(02)	&	13.20(02)	&	13.24(02)	&	12.77(03)	&	REM	\\
60325.55 	&	-3.9 	&	$\cdots$	&	$\cdots$	&	$\cdots$	&	12.97(03)	&	$\cdots$	&	$\cdots$	&	$\cdots$	&	$\cdots$	&	ZTF	\\
60325.90 	&	-3.6 	&	13.81(01)	&	13.10(05)	&	12.89(05)	&	12.97(06)	&	12.89(06)	&	13.31(07)	&	13.09(01)	&	$\cdots$	&	LJT	\\
60326.27 	&	-3.2 	&	$\cdots$	&	$\cdots$	&	$\cdots$	&	13.11(02)	&	12.84(02)	&	13.31(02)	&	13.15(02)	&	$\cdots$	&	REM	\\
60326.96 	&	-2.5 	&	13.80(01)	&	13.07(07)	&	12.86(06)	&	12.95(05)	&	12.84(05)	&	13.27(04)	&	13.24(01)	&	$\cdots$	&	LJT	\\
60327.27 	&	-2.2 	&	$\cdots$	&	$\cdots$	&	$\cdots$	&	13.09(02)	&	12.92(02)	&	13.24(02)	&	$\cdots$	&	12.44(03)	&	REM	\\
60327.94 	&	-1.6 	&	13.71(01)	&	13.03(08)	&	12.81(05)	&	12.91(05)	&	12.79(04)	&	13.31(06)	&	13.20(01)	&	$\cdots$	&	LJT	\\
60328.28 	&	-1.2 	&	$\cdots$	&	$\cdots$	&	$\cdots$	&	13.04(02)	&	12.84(02)	&	13.38(02)	&	13.04(02)	&	12.60(02)	&	REM	\\
60328.94 	&	-0.6 	&	13.74(01)	&	13.05(07)	&	12.80(12)	&	12.92(08)	&	12.79(06)	&	13.34(07)	&	13.19(01)	&	$\cdots$	&	LJT	\\
60329.00 	&	-0.5 	&	13.85(02)	&	12.93(03)	&	12.75(03)	&	12.91(04)	&	12.73(03)	&	13.23(05)	&	$\cdots$	&	$\cdots$	&	Schmidt	\\
60329.86 	&	0.4 	&	13.87(01)	&	13.06(06)	&	12.79(04)	&	12.74(19)	&	12.78(04)	&	13.35(05)	&	13.23(01)	&	$\cdots$	&	LJT	\\
60330.34 	&	0.8 	&	$\cdots$	&	$\cdots$	&	$\cdots$	&	12.99(02)	&	12.78(02)	&	13.33(02)	&	13.23(02)	&	12.51(02)	&	REM	\\
60330.78 	&	1.3 	&	$\cdots$	&	$\cdots$	&	$\cdots$	&	12.96(05)	&	12.76(07)	&	13.32(05)	&	$\cdots$	&	$\cdots$	&	TNT	\\
60331.34 	&	1.8 	&	$\cdots$	&	$\cdots$	&	$\cdots$	&	13.12(02)	&	12.96(02)	&	13.32(02)	&	13.19(02)	&	12.58(02)	&	REM	\\
60331.78 	&	2.3 	&	$\cdots$	&	13.24(21)	&	12.85(11)	&	13.02(16)	&	12.80(05)	&	13.38(12)	&	$\cdots$	&	$\cdots$	&	TNT	\\
60331.95 	&	2.5 	&	13.89(01)	&	13.05(03)	&	12.81(08)	&	12.95(06)	&	12.79(06)	&	13.44(05)	&	13.26(01)	&	$\cdots$	&	LJT	\\
60332.00 	&	2.5 	&	14.03(10)	&	13.03(02)	&	12.78(02)	&	12.95(04)	&	12.72(03)	&	13.33(03)	&	$\cdots$	&	$\cdots$	&	Schmidt	\\
60332.34 	&	2.8 	&	$\cdots$	&	$\cdots$	&	$\cdots$	&	13.07(02)	&	12.85(02)	&	13.37(02)	&	13.13(02)	&	12.47(03)	&	REM	\\
60332.73 	&	3.2 	&	$\cdots$	&	13.22(08)	&	12.83(04)	&	12.95(07)	&	12.82(05)	&	$\cdots$	&	$\cdots$	&	$\cdots$	&	TNT	\\
60333.34 	&	3.8 	&	$\cdots$	&	$\cdots$	&	$\cdots$	&	12.97(02)	&	12.77(02)	&	13.36(02)	&	13.23(02)	&	12.50(03)	&	REM	\\
60333.71 	&	4.2 	&	$\cdots$	&	13.24(07)	&	12.89(05)	&	13.03(06)	&	12.85(05)	&	13.44(04)	&	$\cdots$	&	$\cdots$	&	TNT	\\
60334.34 	&	4.8 	&	$\cdots$	&	$\cdots$	&	$\cdots$	&	$\cdots$	&	$\cdots$	&	13.34(02)	&	13.35(02)	&	12.68(02)	&	REM	\\
60334.70 	&	5.2 	&	$\cdots$	&	13.24(11)	&	12.90(05)	&	13.04(06)	&	12.86(04)	&	13.44(02)	&	$\cdots$	&	$\cdots$	&	TNT	\\
60335.34 	&	5.8 	&	$\cdots$	&	$\cdots$	&	$\cdots$	&	13.06(02)	&	$\cdots$	&	13.41(02)	&	13.14(02)	&	12.86(03)	&	REM	\\
60335.71 	&	6.2 	&	$\cdots$	&	13.52(21)	&	12.96(10)	&	13.19(15)	&	12.94(06)	&	13.54(12)	&	$\cdots$	&	$\cdots$	&	TNT	\\
60335.76 	&	6.3 	&	$\cdots$	&	$\cdots$	&	$\cdots$	&	13.10(07)	&	12.89(04)	&	13.59(03)	&	13.46(01)	&	$\cdots$	&	LJT	\\
60336.34 	&	6.8 	&	$\cdots$	&	$\cdots$	&	$\cdots$	&	13.22(02)	&	12.80(02)	&	13.53(02)	&	13.36(02)	&	12.60(03)	&	REM	\\
60336.82 	&	7.3 	&	$\cdots$	&	13.54(20)	&	13.03(11)	&	13.26(16)	&	13.01(06)	&	13.62(12)	&	$\cdots$	&	$\cdots$	&	TNT	\\
60337.35 	&	7.9 	&	$\cdots$	&	$\cdots$	&	$\cdots$	&	13.04(02)	&	12.99(02)	&	$\cdots$	&	13.23(02)	&	12.81(04)	&	REM	\\
60338.35 	&	8.9 	&	$\cdots$	&	$\cdots$	&	$\cdots$	&	13.41(02)	&	12.92(02)	&	13.55(02)	&	13.41(02)	&	12.60(03)	&	REM	\\
60339.35 	&	9.9 	&	$\cdots$	&	$\cdots$	&	$\cdots$	&	13.20(02)	&	13.03(02)	&	$\cdots$	&	13.34(02)	&	$\cdots$	&	REM	\\
60340.35 	&	10.9 	&	$\cdots$	&	$\cdots$	&	$\cdots$	&	13.41(02)	&	13.25(02)	&	13.64(02)	&	13.52(02)	&	12.74(02)	&	REM	\\
60341.35 	&	11.9 	&	$\cdots$	&	$\cdots$	&	$\cdots$	&	13.40(02)	&	13.29(02)	&	13.81(02)	&	13.31(02)	&	12.86(02)	&	REM	\\
60341.77 	&	12.3 	&	$\cdots$	&	14.02(21)	&	13.31(11)	&	13.62(15)	&	13.39(07)	&	13.99(12)	&	$\cdots$	&	$\cdots$	&	TNT	\\
60342.35 	&	12.9 	&	$\cdots$	&	$\cdots$	&	$\cdots$	&	13.44(02)	&	13.30(02)	&	13.87(02)	&	13.38(02)	&	12.71(02)	&	REM	\\
60343.00 	&	13.5 	&	15.02(09)	&	13.88(03)	&	13.24(05)	&	13.50(05)	&	13.32(16)	&	13.94(06)	&	$\cdots$	&	$\cdots$	&	Schmidt	\\
60343.35 	&	13.9 	&	$\cdots$	&	$\cdots$	&	$\cdots$	&	13.56(02)	&	13.40(02)	&	13.90(02)	&	13.31(02)	&	12.67(02)	&	REM	\\
60343.77 	&	14.3 	&	$\cdots$	&	14.26(21)	&	13.43(11)	&	13.79(16)	&	13.47(05)	&	13.99(12)	&	$\cdots$	&	$\cdots$	&	TNT	\\
60344.35 	&	14.9 	&	$\cdots$	&	$\cdots$	&	$\cdots$	&	13.75(02)	&	13.49(02)	&	13.76(02)	&	13.46(02)	&	$\cdots$	&	REM	\\
60344.78 	&	15.3 	&	15.47(02)	&	14.20(06)	&	13.43(06)	&	13.76(09)	&	13.42(04)	&	14.03(04)	&	13.51(01)	&	$\cdots$	&	LJT	\\
60345.36 	&	15.9 	&	$\cdots$	&	$\cdots$	&	$\cdots$	&	13.82(02)	&	13.40(02)	&	13.92(02)	&	13.67(02)	&	12.74(05)	&	REM	\\
60346.36 	&	16.9 	&	$\cdots$	&	$\cdots$	&	$\cdots$	&	13.82(02)	&	13.50(02)	&	13.90(02)	&	13.50(02)	&	12.46(03)	&	REM	\\
60347.34 	&	17.8 	&	$\cdots$	&	$\cdots$	&	$\cdots$	&	14.03(02)	&	13.44(02)	&	13.79(02)	&	13.77(02)	&	$\cdots$	&	REM	\\
60347.73 	&	18.2 	&	15.86(02)	&	14.53(01)	&	13.71(14)	&	13.99(05)	&	13.50(06)	&	13.99(04)	&	13.42(01)	&	$\cdots$	&	LJT	\\
60347.83 	&	18.3 	&	$\cdots$	&	14.73(19)	&	13.65(10)	&	14.17(15)	&	13.57(06)	&	13.92(11)	&	$\cdots$	&	$\cdots$	&	TNT	\\
60348.80 	&	19.3 	&	$\cdots$	&	14.85(21)	&	13.69(12)	&	14.27(18)	&	13.58(08)	&	13.88(15)	&	$\cdots$	&	$\cdots$	&	TNT	\\
60349.37 	&	19.9 	&	$\cdots$	&	$\cdots$	&	$\cdots$	&	14.18(02)	&	13.59(02)	&	13.81(02)	&	13.32(02)	&	12.46(03)	&	REM	\\
60350.74 	&	21.2 	&	15.91(02)	&	14.85(24)	&	13.70(10)	&	14.22(09)	&	13.52(03)	&	13.81(13)	&	13.29(01)	&	$\cdots$	&	LJT	\\
60351.37 	&	21.9 	&	$\cdots$	&	$\cdots$	&	$\cdots$	&	14.28(03)	&	$\cdots$	&	$\cdots$	&	$\cdots$	&	$\cdots$	&	ZTF	\\
60351.39 	&	21.9 	&	$\cdots$	&	$\cdots$	&	$\cdots$	&	14.20(02)	&	13.56(02)	&	13.65(02)	&	13.39(02)	&	12.58(02)	&	REM	\\
60355.30 	&	25.8 	&	$\cdots$	&	$\cdots$	&	$\cdots$	&	14.67(02)	&	13.71(02)	&	13.68(02)	&	13.50(02)	&	12.38(02)	&	REM	\\
60355.41 	&	25.9 	&	$\cdots$	&	$\cdots$	&	$\cdots$	&	14.64(04)	&	$\cdots$	&	$\cdots$	&	$\cdots$	&	$\cdots$	&	ZTF	\\
60356.74 	&	27.2 	&	16.65(02)	&	15.40(08)	&	14.08(09)	&	14.79(07)	&	13.74(07)	&	13.84(08)	&	13.19(01)	&	$\cdots$	&	LJT	\\
60357.30 	&	27.8 	&	$\cdots$	&	$\cdots$	&	$\cdots$	&	14.78(02)	&	13.75(02)	&	13.72(02)	&	13.31(02)	&	12.37(02)	&	REM	\\
60358.81 	&	29.3 	&	16.68(03)	&	15.60(11)	&	14.25(08)	&	14.98(08)	&	13.93(14)	&	13.93(10)	&	13.40(01)	&	$\cdots$	&	LJT	\\
60359.36 	&	29.9 	&	$\cdots$	&	$\cdots$	&	$\cdots$	&	15.03(02)	&	$\cdots$	&	13.72(02)	&	13.49(02)	&	12.67(02)	&	REM	\\
60362.76 	&	33.3 	&	$\cdots$	&	15.91(20)	&	14.47(10)	&	15.36(13)	&	14.14(05)	&	14.10(11)	&	$\cdots$	&	$\cdots$	&	TNT	\\
60364.75 	&	35.2 	&	$\cdots$	&	16.21(23)	&	14.65(13)	&	15.57(17)	&	14.29(07)	&	14.28(11)	&	$\cdots$	&	$\cdots$	&	TNT	\\
60365.27 	&	35.8 	&	$\cdots$	&	$\cdots$	&	$\cdots$	&	15.24(02)	&	14.08(02)	&	14.00(02)	&	13.99(02)	&	12.81(03)	&	REM	\\
60368.43 	&	38.9 	&	$\cdots$	&	$\cdots$	&	$\cdots$	&	15.43(04)	&	14.44(03)	&	$\cdots$	&	$\cdots$	&	$\cdots$	&	ZTF	\\
60369.73 	&	40.2 	&	$\cdots$	&	16.24(19)	&	14.81(11)	&	15.67(13)	&	14.54(06)	&	14.47(09)	&	$\cdots$	&	$\cdots$	&	TNT	\\
60370.70 	&	41.2 	&	$\cdots$	&	16.22(10)	&	14.82(05)	&	15.65(04)	&	14.58(02)	&	14.50(03)	&	$\cdots$	&	$\cdots$	&	TNT	\\
60371.77 	&	42.3 	&	$\cdots$	&	16.17(04)	&	14.82(03)	&	15.64(05)	&	14.59(05)	&	14.53(05)	&	$\cdots$	&	$\cdots$	&	TNT	\\
60372.73 	&	43.2 	&	$\cdots$	&	16.29(06)	&	14.86(03)	&	15.65(03)	&	14.65(03)	&	14.59(03)	&	$\cdots$	&	$\cdots$	&	TNT	\\
60374.42 	&	44.9 	&	$\cdots$	&	$\cdots$	&	$\cdots$	&	15.55(04)	&	$\cdots$	&	$\cdots$	&	$\cdots$	&	$\cdots$	&	ZTF	\\
60375.68 	&	46.2 	&	$\cdots$	&	16.50(25)	&	15.04(15)	&	15.80(19)	&	14.78(08)	&	14.76(15)	&	$\cdots$	&	$\cdots$	&	TNT	\\
60377.71 	&	48.2 	&	$\cdots$	&	16.41(15)	&	15.00(09)	&	15.81(13)	&	14.82(05)	&	14.82(09)	&	$\cdots$	&	$\cdots$	&	TNT	\\
60380.28 	&	50.8 	&	$\cdots$	&	$\cdots$	&	$\cdots$	&	15.70(03)	&	$\cdots$	&	$\cdots$	&	$\cdots$	&	$\cdots$	&	ZTF	\\
60380.78 	&	51.3 	&	17.48(03)	&	$\cdots$	&	$\cdots$	&	15.73(09)	&	14.95(07)	&	15.14(09)	&	14.70(01)	&	$\cdots$	&	LJT	\\
60389.60 	&	60.1 	&	$\cdots$	&	16.45(08)	&	15.31(04)	&	15.94(03)	&	15.20(03)	&	15.23(04)	&	$\cdots$	&	$\cdots$	&	TNT	\\
60389.85 	&	60.3 	&	$\cdots$	&	$\cdots$	&	15.30(05)	&	15.80(09)	&	15.19(06)	&	15.35(07)	&	15.21(02)	&	$\cdots$	&	LJT	\\
60390.42 	&	60.9 	&	$\cdots$	&	$\cdots$	&	$\cdots$	&	15.82(04)	&	$\cdots$	&	$\cdots$	&	$\cdots$	&	$\cdots$	&	ZTF	\\
60397.41 	&	67.9 	&	$\cdots$	&	$\cdots$	&	$\cdots$	&	$\cdots$	&	15.49(04)	&	$\cdots$	&	$\cdots$	&	$\cdots$	&	ZTF	\\
60401.76 	&	72.3 	&	$\cdots$	&	16.50(07)	&	15.66(06)	&	$\cdots$	&	15.55(02)	&	15.84(07)	&	$\cdots$	&	$\cdots$	&	LJT	\\
60403.32 	&	73.8 	&	$\cdots$	&	$\cdots$	&	$\cdots$	&	16.03(04)	&	$\cdots$	&	$\cdots$	&	$\cdots$	&	$\cdots$	&	ZTF	\\
60406.60 	&	77.1 	&	$\cdots$	&	16.66(10)	&	15.82(08)	&	16.22(04)	&	15.74(01)	&	15.83(02)	&	$\cdots$	&	$\cdots$	&	TNT	\\
60407.32 	&	77.8 	&	$\cdots$	&	$\cdots$	&	$\cdots$	&	16.06(04)	&	$\cdots$	&	$\cdots$	&	$\cdots$	&	$\cdots$	&	ZTF	\\
60408.63 	&	79.1 	&	$\cdots$	&	16.84(10)	&	15.83(09)	&	16.25(07)	&	15.86(07)	&	15.94(08)	&	$\cdots$	&	$\cdots$	&	TNT	\\
60408.74 	&	79.2 	&	$\cdots$	&	16.59(08)	&	15.83(06)	&	16.17(07)	&	15.83(07)	&	16.11(08)	&	15.95(04)	&	$\cdots$	&	LJT	\\
60409.23 	&	79.7 	&	$\cdots$	&	$\cdots$	&	$\cdots$	&	16.12(03)	&	15.82(02)	&	$\cdots$	&	$\cdots$	&	$\cdots$	&	ZTF	\\
60410.27 	&	80.8 	&	$\cdots$	&	$\cdots$	&	$\cdots$	&	16.14(02)	&	15.86(02)	&	$\cdots$	&	$\cdots$	&	$\cdots$	&	ZTF	\\
60411.33 	&	81.8 	&	$\cdots$	&	$\cdots$	&	$\cdots$	&	16.15(02)	&	15.95(03)	&	$\cdots$	&	$\cdots$	&	$\cdots$	&	ZTF	\\
60412.33 	&	82.8 	&	$\cdots$	&	$\cdots$	&	$\cdots$	&	16.17(04)	&	$\cdots$	&	$\cdots$	&	$\cdots$	&	$\cdots$	&	ZTF	\\
60413.00 	&	83.5 	&	$\cdots$	&	$\cdots$	&	$\cdots$	&	$\cdots$	&	15.86(03)	&	16.07(03)	&	$\cdots$	&	$\cdots$	&	Schmidt	\\
60416.30 	&	86.8 	&	$\cdots$	&	$\cdots$	&	$\cdots$	&	16.21(04)	&	$\cdots$	&	$\cdots$	&	$\cdots$	&	$\cdots$	&	ZTF	\\
60417.55 	&	88.1 	&	$\cdots$	&	16.75(12)	&	16.07(07)	&	16.35(06)	&	16.09(06)	&	16.16(08)	&	$\cdots$	&	$\cdots$	&	TNT	\\
60418.23 	&	88.7 	&	$\cdots$	&	$\cdots$	&	$\cdots$	&	16.22(04)	&	16.12(04)	&	$\cdots$	&	$\cdots$	&	$\cdots$	&	ZTF	\\
60429.29 	&	99.8 	&	$\cdots$	&	$\cdots$	&	$\cdots$	&	16.43(05)	&	$\cdots$	&	$\cdots$	&	$\cdots$	&	$\cdots$	&	ZTF	\\
60431.29 	&	101.8 	&	$\cdots$	&	$\cdots$	&	$\cdots$	&	16.44(05)	&	$\cdots$	&	$\cdots$	&	$\cdots$	&	$\cdots$	&	ZTF	\\
60434.30 	&	104.8 	&	$\cdots$	&	$\cdots$	&	$\cdots$	&	16.50(04)	&	$\cdots$	&	$\cdots$	&	$\cdots$	&	$\cdots$	&	ZTF	\\
60434.74 	&	105.2 	&	$\cdots$	&	16.94(05)	&	16.41(06)	&	16.55(05)	&	16.57(05)	&	16.88(07)	&	16.79(04)	&	$\cdots$	&	LJT	\\
60438.30 	&	108.8 	&	$\cdots$	&	$\cdots$	&	$\cdots$	&	16.55(04)	&	$\cdots$	&	$\cdots$	&	$\cdots$	&	$\cdots$	&	ZTF	\\
60438.76 	&	109.3 	&	$\cdots$	&	$\cdots$	&	16.45(07)	&	$\cdots$	&	16.68(05)	&	16.96(11)	&	$\cdots$	&	$\cdots$	&	LJT	\\
60440.31 	&	110.8 	&	$\cdots$	&	$\cdots$	&	$\cdots$	&	16.57(05)	&	$\cdots$	&	$\cdots$	&	$\cdots$	&	$\cdots$	&	ZTF	\\
60440.69 	&	111.2 	&	$\cdots$	&	$\cdots$	&	16.56(07)	&	$\cdots$	&	$\cdots$	&	$\cdots$	&	$\cdots$	&	$\cdots$	&	LJT	\\
60442.19 	&	112.7 	&	$\cdots$	&	$\cdots$	&	$\cdots$	&	16.58(04)	&	$\cdots$	&	$\cdots$	&	$\cdots$	&	$\cdots$	&	ZTF	\\
60443.73 	&	114.2 	&	$\cdots$	&	17.08(07)	&	16.58(07)	&	16.68(06)	&	16.81(03)	&	17.13(06)	&	$\cdots$	&	$\cdots$	&	LJT	\\
60444.21 	&	114.7 	&	$\cdots$	&	$\cdots$	&	$\cdots$	&	16.63(04)	&	$\cdots$	&	$\cdots$	&	$\cdots$	&	$\cdots$	&	ZTF	\\
60446.21 	&	116.7 	&	$\cdots$	&	$\cdots$	&	$\cdots$	&	16.65(05)	&	16.91(05)	&	$\cdots$	&	$\cdots$	&	$\cdots$	&	ZTF	\\
60450.24 	&	120.7 	&	$\cdots$	&	$\cdots$	&	$\cdots$	&	$\cdots$	&	17.00(04)	&	$\cdots$	&	$\cdots$	&	$\cdots$	&	ZTF	\\
\enddata
\tablecomments{Uncertainties are enclosed in parentheses and are $1\sigma$, in units of 0.01 mag. }
$^a${Phase relative to the day of $B$-band maximum, MJD = 60329.5.}\\
\end{deluxetable*}

\begin{deluxetable*}{lccccccc}
\tablecaption{Mephisto \textit{uvgriz} Photometry of SN 2024gy\label{app_a:M_photo}}
\tabletypesize{\scriptsize}
\tablewidth{0pt}
\tablehead{
\colhead{MJD} & \colhead{Phase\tablenotemark{a}} & \colhead{\textit{u}} & \colhead{\textit{v}} & \colhead{\textit{g}}& \colhead{\textit{r}}& \colhead{\textit{i}}& \colhead{\textit{z}}\\
\colhead{} & \colhead{days} & \colhead{mag} & \colhead{mag} & \colhead{mag}& \colhead{mag}& \colhead{mag}& \colhead{mag}
}
\startdata
60314.87 	&	-14.6 	&	19.32(09)	&	16.57(01)	&	15.58(01)	&	15.42(01)	&	$\cdots$	&	$\cdots$	\\
60316.88 	&	-12.6 	&	16.75(01)	&	15.06(01)	&	14.60(01)	&	14.46(01)	&	$\cdots$	&	$\cdots$	\\
60317.89 	&	-11.6 	&	16.04(01)	&	14.58(01)	&	14.23(01)	&	14.06(01)	&	$\cdots$	&	$\cdots$	\\
60318.87 	&	-10.6 	&	15.01(01)	&	14.24(01)	&	13.97(01)	&	13.75(01)	&	$\cdots$	&	$\cdots$	\\
60319.80 	&	-9.7 	&	14.77(01)	&	13.97(01)	&	13.74(01)	&	13.54(01)	&	$\cdots$	&	$\cdots$	\\
60320.94 	&	-8.6 	&	14.26(01)	&	13.74(01)	&	13.51(01)	&	13.31(01)	&	13.58(01)	&	13.59(01)	\\
60321.95 	&	-7.6 	&	14.00(01)	&	13.62(01)	&	13.30(01)	&	13.15(01)	&	13.47(01)	&	13.49(01)	\\
60322.88 	&	-6.6 	&	13.51(01)	&	13.51(01)	&	13.18(01)	&	13.05(01)	&	13.39(01)	&	13.41(01)	\\
60324.90 	&	-4.6 	&	13.79(01)	&	13.35(01)	&	12.99(01)	&	12.90(01)	&	13.30(01)	&	13.35(01)	\\
60325.86 	&	-3.6 	&	13.65(01)	&	13.28(01)	&	12.94(01)	&	12.73(01)	&	13.26(01)	&	13.33(01)	\\
60326.86 	&	-2.6 	&	13.64(01)	&	13.22(01)	&	12.91(01)	&	12.77(01)	&	13.23(01)	&	13.34(01)	\\
60327.88 	&	-1.6 	&	13.65(01)	&	13.25(01)	&	12.87(01)	&	12.81(01)	&	13.21(01)	&	13.36(01)	\\
60328.82 	&	-0.7 	&	13.64(01)	&	13.26(01)	&	12.86(01)	&	12.76(01)	&	13.17(01)	&	13.39(01)	\\
60329.84 	&	0.3 	&	13.78(01)	&	13.27(01)	&	12.84(01)	&	12.72(01)	&	13.21(01)	&	13.47(01)	\\
60330.94 	&	1.4 	&	13.84(01)	&	13.29(01)	&	12.83(01)	&	12.69(01)	&	13.23(01)	&	13.53(01)	\\
60331.80 	&	2.3 	&	13.88(01)	&	$\cdots$	&	12.86(01)	&	$\cdots$	&	13.21(01)	&	$\cdots$	\\
60335.79 	&	6.3 	&	14.69(02)	&	13.69(01)	&	13.01(01)	&	12.82(01)	&	13.45(01)	&	13.79(01)	\\
60341.91 	&	12.4 	&	$\cdots$	&	14.21(01)	&	13.48(01)	&	13.24(01)	&	$\cdots$	&	13.85(01)	\\
60347.89 	&	18.4 	&	16.22(01)	&	14.94(01)	&	13.70(01)	&	13.40(01)	&	13.60(01)	&	13.63(01)	\\
60348.80 	&	19.3 	&	$\cdots$	&	$\cdots$	&	$\cdots$	&	13.32(01)	&	$\cdots$	&	$\cdots$	\\
60349.82 	&	20.3 	&	16.51(01)	&	15.23(01)	&	13.80(01)	&	13.48(01)	&	13.60(01)	&	13.59(01)	\\
60350.73 	&	21.2 	&	16.52(02)	&	15.37(01)	&	13.80(01)	&	13.51(01)	&	13.58(01)	&	13.58(01)	\\
60351.82 	&	22.3 	&	16.63(02)	&	15.43(01)	&	13.84(01)	&	13.43(01)	&	13.56(01)	&	13.54(01)	\\
60354.91 	&	25.4 	&	16.90(02)	&	15.72(01)	&	14.00(01)	&	13.59(01)	&	13.54(01)	&	13.45(01)	\\
60355.83 	&	26.3 	&	17.01(02)	&	15.79(01)	&	14.06(01)	&	13.63(01)	&	13.57(01)	&	13.46(01)	\\
60356.78 	&	27.3 	&	$\cdots$	&	15.87(01)	&	14.12(01)	&	13.70(01)	&	$\cdots$	&	13.41(01)	\\
60357.80 	&	28.3 	&	17.15(02)	&	16.00(01)	&	14.15(01)	&	13.77(01)	&	13.54(01)	&	13.44(01)	\\
60359.81 	&	30.3 	&	17.32(02)	&	16.11(01)	&	14.32(01)	&	13.84(01)	&	13.67(01)	&	13.51(01)	\\
60374.75 	&	45.2 	&	17.76(02)	&	16.56(01)	&	15.02(01)	&	14.66(01)	&	14.69(01)	&	14.41(01)	\\
60385.73 	&	56.2 	&	18.03(04)	&	16.74(01)	&	15.40(01)	&	14.99(01)	&	15.30(01)	&	14.93(01)	\\
60388.82 	&	59.3 	&	18.27(06)	&	16.84(02)	&	15.51(01)	&	15.07(01)	&	15.47(01)	&	15.11(02)	\\
60403.71 	&	74.2 	&	$\cdots$	&	17.08(02)	&	15.90(06)	&	15.49(01)	&	16.28(22)	&	15.71(02)	\\
60410.58 	&	81.1 	&	18.57(04)	&	17.23(01)	&	16.07(01)	&	15.77(01)	&	16.46(02)	&	15.93(02)	\\
60418.56 	&	89.1 	&	18.71(09)	&	17.44(03)	&	16.22(02)	&	15.88(02)	&	16.85(02)	&	16.21(03)	\\
60426.56 	&	97.1 	&	$\cdots$	&	17.60(02)	&	$\cdots$	&	16.26(01)	&	$\cdots$	&	16.35(03)	\\
60429.54 	&	100.0 	&	18.94(06)	&	$\cdots$	&	16.50(02)	&	16.28(02)	&	17.13(04)	&	$\cdots$	\\
60434.66 	&	105.2 	&	$\cdots$	&	17.84(03)	&	$\cdots$	&	16.45(02)	&	$\cdots$	&	16.70(04)	\\
60455.57 	&	126.1 	&	19.07(08)	&	18.36(03)	&	17.01(01)	&	17.11(01)	&	17.98(02)	&	17.08(03)	\\
60629.93 	&	300.4 	&	$\cdots$	&	$\cdots$	&	19.82(12)	&	21.13(25)	&	$\cdots$	&	$\cdots$	\\
60645.93 	&	316.4 	&	$\cdots$	&	$\cdots$	&	20.13(07)	&	21.14(29)	&	21.09(19)	&	19.53(37)	\\
60761.75 	&	432.3 	&	$\cdots$	&	$\cdots$	&	$\cdots$	&	22.89(57)	&	$\cdots$	&	$\cdots$	\\
\enddata
\tablecomments{Uncertainties are enclosed in parentheses and are $1\sigma$, in units of 0.01 mag. }
\tablenotetext{a}{Phase relative to the day of $B$-band maximum, MJD = 60329.5.}
\end{deluxetable*}
\begin{deluxetable*}{cccccc}
\tablecaption{Journal of Spectroscopic Observations\label{app_a:spec}}
\tabletypesize{\small}
\centering
\tablehead{
\colhead{MJD} & \colhead{Phase\tablenotemark{a} } & \colhead{Range(\AA)} & \colhead{R\tablenotemark{*}} & \colhead{Airmass} & \colhead{Telescope+Inst.}
}
\startdata
60314.81 	&	-14.7 	&	3616-8926  &  320   &  1.36	&	LJT+YFOSC	\\
60314.89 	&	-14.6 	&	3776-8915  &    300    &  1.16	&	XLT+BFOSC	\\
60314.91 	&	-14.6 	&	3616-8926  &  320   &  1.04	&	LJT+YFOSC	\\
60315.45 	&	-14.0 	&	3417-9000  &  940/1220  &  1.33	&	Shane+Kast	\\
60315.80 	&	-13.7 	&	3773-8918  &    300    &  1.18	&	XLT+BFOSC	\\
60315.92 	&	-13.6 	&	3454-8925  &  315   &  1.03	&	LJT+YFOSC	\\
60316.88 	&	-12.6 	&	3454-8925  &  320   &  1.08	&	LJT+YFOSC	\\
60317.44 	&	-12.1 	&	3504-11450 &  1250/950  &  1.35	&	Shane+Kast	\\
60317.88 	&	-11.6 	&	3456-8924  &  315   &  1.07	&	LJT+YFOSC	\\
60318.92    &	-10.6 	&	4900-7589  & 3500   &  1.03	&	LJT+YFOSC	\\
60319.89 	&	-9.6 	&	3454-8927   &  320  &  1.04	&	LJT+YFOSC	\\
60320.94 	&	-8.6 	&	3450-8925   &  320  &  1.04	&	LJT+YFOSC	\\
60321.43 	&	-8.1 	&	3638-10670  &  970/690 &  1.35	&	Shane+Kast	\\
60323.84    &	-5.7 	&	4000-7594   & 3500  &  1.12	&	LJT+YFOSC	\\
60324.59 	&	-4.9 	&	3640-8580   &  980/1680 &  1.15	&	Shane+Kast	\\
60325.88 	&	-3.6 	&	3457-8925   &  335  &  1.04	&	LJT+YFOSC	\\
60327.59 	&	-1.9 	&	3638-10748  &  970/690 &  1.19	&	Shane+Kast	\\
60327.90    &	-1.6 	&	4000-7592  &  3500  &  1.04	&	LJT+YFOSC	\\
60328.59 	&	-0.9 	&	3636-8736   & 1940/2360  &  1.20	&	Shane+Kast	\\
60329.85 	&	0.3 	&	3618-8926   &  300  &  1.08 &	LJT+YFOSC	\\
60331.87 	&	2.4 	&	3776-8919   &  260  &  1.20	&	XLT+BFOSC	\\
60342.88 	&	13.4 	&	3779-8917   &   270   &  1.36	&	XLT+BFOSC	\\
60347.00 	&	17.5 	&	3147-10252  &  940/920 &  1.27	&	Keck I+LRIS	\\
60348.73 	&	19.2 	&	3615-8927   &  320 &  1.31 &	LJT+YFOSC	\\
60350.70    &   21.3    &   4592-7594   & 3500  &  1.45 &   LJT+YFOSC	\\
60356.77 	&	27.3 	&	3775-8912   &   310   &  1.13	&	XLT+BFOSC	\\
60369.80 	&	40.3 	&	3780-8914   &    300   &  1.28	&	XLT+BFOSC	\\
60386.46    &   56.0    &   3624-10748  &  970/690 &  1.30  &   Shane+Kast	\\
60434.23    &   104.7   &   3142-10249  &  940/920 &  1.19 &   Keck I+LRIS	\\
60443.70 	&	114.2 	&	3613-8919  &   300  &  1.37	&	LJT+YFOSC	\\
60461.24    &   131.7   &   3636-10740  &  970/690 &  1.24  &   Shane+Kast	\\
60467.21    &   137.7   &   3636-10760  & 970/690 &  1.19  &   Shane+Kast	\\
60474.30    &   144.8   &   3626-10756  &  970/690 &  2.14  &   Shane+Kast	\\
60648.49    &   320.0   &   3610-10756  &  970/690 &  1.71  &   Shane+Kast	\\
60733.59 	&	404.1 	&	3147-10258 &  940/930  &  1.19	&	Keck I+LRIS	\\
\enddata
\tablenotetext{a}{Phase relative to the day of $B$-band maximum, MJD = 60329.5.}
\tablenotetext{*}{The resoving power for the blue and red channels of the Kast/LRIS spectra is provided by the individual cameras in each spectrograph arm.}
\end{deluxetable*}

\end{document}